%% file: arXiv-cloud-main.tex
\documentclass[journal]{vgtc-clear}            
\onlineid{1779}

\vgtccategory{Research}
\vgtcpapertype{Application}

\graphicspath{{./figs/}} 

\usepackage{tabu}         
\usepackage{booktabs} 
\usepackage{amsmath}
\usepackage{amsfonts}

\newcommand{\ie}{{i.e.}}

\newcommand{\para}[1]{\vspace{0mm}\noindent{\textbf{#1}}}

\newcommand{\mm}[1]{\ifmmode{#1}\else{\mbox{\(#1\)}}\fi}

\newcommand{\pFGW}{\textsf{pFGW}}
\newcommand{\tobac}{\textsf{tobac}}
\newcommand{\PyFT}{\textsf{PyFLEXTRKR}}
\newcommand{\LWM}{\textsf{LWM}}

\newcommand{\MTW}{\textsf{MTW}}

\newcommand{\onevec}[1]{\mm{\mathbf{1}_{#1}}}
\newcommand{\M}{\mathbb{M}}
\newcommand{\R}{\mathbb{R}}
\newcommand{\Ccal}{\mm{\mathcal{C}}}

\newcommand{\DM}{\textbf{Marine Cloud}}
\newcommand{\DL}{\textbf{Land Cloud}}

\newcommand{\CODm}{\textrm{mean\,COD}}

\title{Tracking Low-Level Cloud Systems with Topology}

\author{
  \myauthor{Mingzhe Li}, 
  \myauthor{Dwaipayan Chatterjee},
  \myauthor{Franziska Glassmeier},
  \myauthor{Fabian Senf},
  \myauthor{Bei Wang}
}

\authorfooter{
\item Mingzhe Li is with the University of Utah.
E-mail: mingzhe.li@utah.edu.
\item Dwaipayan Chatterjee is with the Karlsruhe Institute of Technology. E-mail: dwaipayan.chatterjee@kit.edu.
\item Franziska Glassmeier is with the TU Delft. 
    E-mail: f.glassmeier@tudelft.nl.
\item Fabian Senf is with the Leibniz Institute for Troposheric Research. 
    E-mail: senf@tropos.de. 
\item Bei Wang is with the University of Utah. 
    E-mail: beiwang@sci.utah.edu. 
}

\abstract{
\input{sec-abstract}

}


\teaser{
    \includegraphics[width=0.9\linewidth]{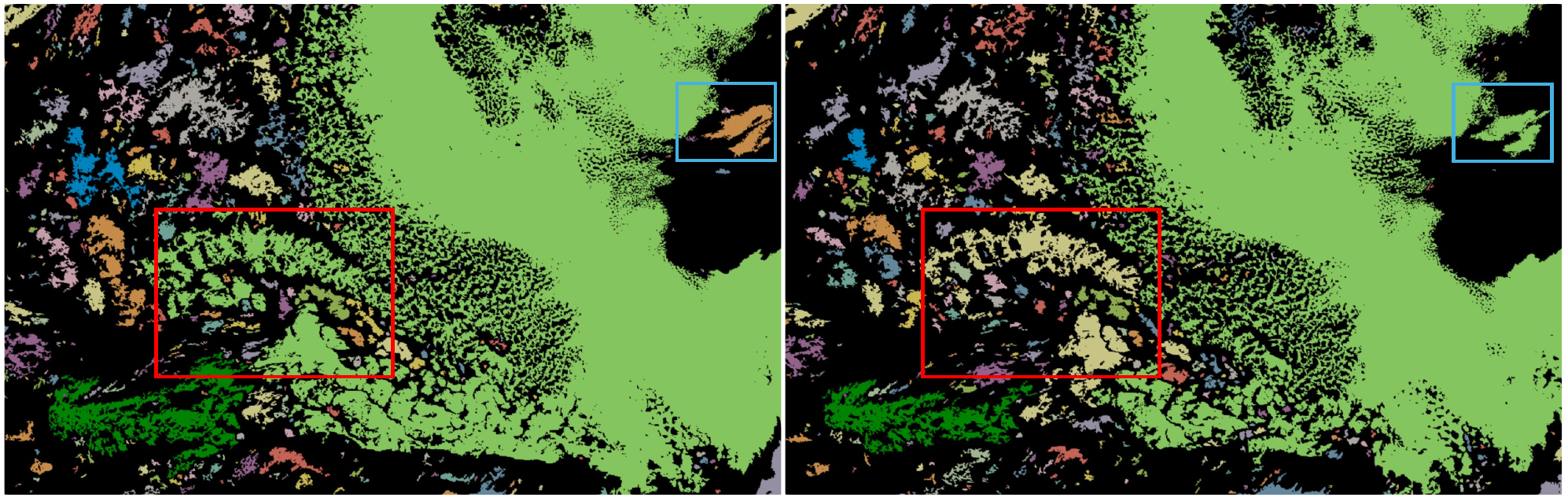}
    \centering
    \vspace{-3mm}
    \caption{Topology-driven cloud tracking results for a marine stratocumulus cloud dataset over the ocean west of Africa. The two images show the detected cloud systems at 09:00 and 10:00 UTC on Aug 1, 2023, respectively. Color encodings between the images indicate the correspondences between cloud systems over a time span of an hour. We can observe the evolution of these cloud systems as they appear, disappear, merge, and split. We use red and cyan boxes to highlight a splitting event and a merging event, respectively.}
    \label{fig:teaser}
}

\begin{document}
\maketitle

\input{sec-introduction}
\input{sec-cloud-science}
\input{sec-related-work}

\input{sec-background}

\input{sec-method}

\input{sec-results}

\input{sec-evaluation}

\input{sec-discussion}


\bibliographystyle{abbrv-doi-hyperref}
\bibliography{refs-cloud, refs-ot}

\clearpage
\newpage
\appendix 
\input{sec-cloud-data}
\input{sec-implementation}

\input{sec-parameters}
\input{sec-topotools-comparison}


\end{document}

%% file: sec-abstract.tex
Low-level clouds are ubiquitous in Earth's atmosphere, playing a crucial role in transporting heat, moisture, and momentum across the planet. Their evolution and interaction with other atmospheric components, such as aerosols, are essential to understanding the climate system and its sensitivity to anthropogenic influences.      
Advanced high-resolution geostationary satellites now resolve cloud systems with greater accuracy, establishing cloud tracking as a vital research area for studying their spatiotemporal dynamics. It enables disentangling advective and convective components driving cloud evolution. This, in turn, provides deeper insights into the structure and lifecycle of low-level cloud systems and the atmospheric processes they govern.
In this paper, we propose a novel framework for tracking cloud systems using topology-driven techniques based on optimal transport.  We first obtain a set of anchor points for the cloud systems based on the merge tree of the cloud optical depth field. We then apply topology-driven probabilistic feature tracking of these anchor points to guide the tracking of cloud systems. We demonstrate the utility of our framework by tracking clouds over the ocean and land to test for systematic differences in the two physically distinct settings. We further evaluate our framework through case studies and statistical analyses, comparing it against two leading cloud tracking tools and two topology-based general-purpose tracking tools. The results demonstrate that incorporating system-based tracking improves the ability to capture the evolution of low-level clouds. Our framework paves the way for detailed low-level cloud characterization studies using satellite data records.

%% file: sec-introduction.tex
\section{Introduction}
\label{sec:introduction}

Low-level clouds (i.e.,~clouds with a base below 6,500 ft and limited vertical extend) are ubiquitous in Earth’s atmosphere, playing a crucial role in transporting heat, moisture, and momentum across the planet. 
Understanding the structure and evolution of low-level clouds is key to improving our knowledge of atmospheric processes, including their effects on the radiation budget~\cite{trenberth_2009} and precipitation patterns~\cite{vogel_2021}. 
Low-level clouds such as \emph{shallow cumulus} clouds  (colloquially known as \emph{fair weather clouds} with their cotton-ball-like appearance) and \emph{stratocumulus} clouds (gray or whitish cloud decks patterned by dark, cloud-free lines or cells~\cite{WMO-No.407}) significantly affect the Earth's radiative budget because they form cloud fields that can stretch over hundreds of kilometers. Due to their fine-scale structure, however, their accurate representation in climate models remains a significant challenge and a dominant contribution to the uncertainty in climate projections~\cite{Bellouin2020,Sherwood2020}.
Accurately describing these clouds in space as well as time is a prerequisite for their reliable representation in climate models. For example, Dorrestijn et al.~\cite{dorrestijn2013stochastic} evaluated how shallow cumulus convection can be parameterized in climate models, emphasizing the importance of cloud characterization and tracking~\cite{clement_2009, freeman_2024}. 

\emph{Convective} and \emph{advective} mechanisms play a crucial role in shaping the cloud lifecycle. Convective motion, driven by buoyancy and atmospheric instability, governs local-scale vertical movements, including updrafts and downdrafts in convective systems. These dynamics influence cloud growth, dissipation, and structural changes. In contrast, advection refers to the large-scale horizontal transport of clouds by prevailing winds, guided by atmospheric circulation patterns. Cloud tracking, therefore, helps differentiate convective and advective processes~\cite{bley_2016, senf_2015}, enabling studies on cloud dynamics \cite{Hagos_2019}, lifecycles \cite{Seelig_2021}, microphysical processes \cite{kocher_2022}, and the evolution of cloud patterns. What makes it interesting and different from classical computer vision problems is that clouds are a representation of a continuous process (such as condensation, evaporation, split, and merge) and cannot be treated as a fixed object \cite{dawe_austin_2012}. 
Clouds occur in moist turbulent flow, which results in fractal characteristics of cloud boundaries which in turn makes individual clouds difficult to identify and track. 
Recent progress in tracking methods, driven by growing satellite data records, has made it possible to study cloud characteristics in greater detail and improve our understanding of their behavior.

In this paper, we present a novel topology-driven framework for tracking low-level clouds. 
We model low-level clouds as cloud systems that may consist of multiple cloud objects that are geometrically close, and use probabilistic feature tracking based on optimal transport to track them in a time-varying setting. 
Our contributions are as follows:
\begin{itemize}[noitemsep,leftmargin=*]
\item We present a new framework to track cloud systems using time-varying satellite image data. We first obtain a set of anchor points for the cloud systems based on the merge tree of the cloud optical depth field (description in \cref{sec:cloud-science}). We then apply merge-tree-based feature tracking of the anchor points to guide the tracking of cloud systems. 
\item We showcase the utility of our framework by tracking cloud systems using satellite data over both the ocean and land, as cloud-tracking challenges could differ in the two physically distinct regimes. 
\item We further compare our framework with two leading  cloud tracking tools and two topology-based general-purpose tracking tools via visualizations and statistical evaluations. 
\end{itemize}
The source code and datasets will be made open-source upon acceptance of the paper. 

%% file: sec-cloud-science.tex
\section{Related Work: Cloud Science and Cloud Tracking}
\label{sec:cloud-science}

Clouds form and evolve across various spatiotemporal scales within Earth's atmosphere. Their evolution is governed by the prevailing large-scale atmospheric conditions, intricate interplays of intermediate-scale processes such as locally generated thermodynamic fluxes, and microscopic processes on the cloud-formation scale \cite{klein_2017, naud_2023}. The development of turbulent structures in clouds is influenced by such fluxes and other factors like shear and surface heterogeneity. Consequently, clouds interact with the surrounding turbulent field, altering the dynamics of the boundary layer and coupling it with the free troposphere. Such interactions influence the development of their convective states and facilitate coupling with neighboring clouds \cite{Rick_2014}. 
Additionally, clouds are critical in transporting temperature, moisture, and momentum, significantly contributing to atmospheric circulation \cite{grant_and_lock_2004}. As such, they are integral to regulating Earth's energy balance and water cycle. 

Land and ocean low-level convection exhibit distinct characteristics \cite{Eastman_2014}. Overland convection is driven by rapid heating, variable moisture, and pronounced diurnal cycles, peaking in the late afternoon due to maximum surface heating. Land surface topography often triggers intense but short-lived events. In contrast, oceanic convection~\cite{Seelig_2021} is more organized and sustained, shaped by the ocean's high heat capacity, abundant moisture, uniform temperatures, and large-scale atmospheric circulation.

Low-level clouds in lower latitudes are primarily composed of liquid droplets. These clouds are typically warmer and effective at re-emitting absorbed radiation\cite{bony_2004}. Their response to the warming of the Earth system introduces a key uncertainty in climate projections. 

Advanced Geostationary Satellites (GS), such as Meteosat Third Generation (MTG, \cite{Holmlund_2021}) and GOES-16's Advanced Baseline Imager (ABI, \cite{Schmit_2005}), provide high spatial and temporal resolution. However, they still struggle to resolve the fine-scale structure of many low cloud systems \cite{han_1994, henken_2011}, which often span scales of hundreds of meters. Despite this limitation, treating the aggregation of these individual cloud objects—spanning hundreds of kilometers—as spatial distributions enables adequate resolution to capture their variability \cite{Chatterjee_2024, schulz_characterization_2021, stevens_2021}.

Cloud tracking begins with the identification of individual clouds in the dataset, a process that depends on the type of cloud and the specific scientific objectives of the study. For instance, studies focusing on deep convective cells often rely on physical threshold-based approaches, such as brightness temperature \cite{Fiolleau2013, senf_2015,SenfKlockeBrueck2018} or radar reflectivity \cite{Rosenfeld_1987, valliappa_2010}. In the case of mixed-phase clouds, methods typically involve a combination of cloud mask (cloudy or non-cloudy pixel) and cloud optical depth~\cite{coopman_2019}. Cloud optical depth (COD) quantifies the extent to which the cloud attenuates light primarily due to the scattering and absorption by cloud droplets. It is governed by cloud geometric thickness, cloud water mass, droplet concentration, and particle size distribution \cite{Nakajima1990}. For shallow low-level clouds, cloud identification methods vary: \cite{Seelig_2021} employs a cloud mask, while \cite{kowch_2022} applies a reflectivity threshold. However, shallow cumulus clouds, being inherently broken and scattered in nature, pose additional challenges. The resolution of the cloud mask, such as that provided by the CLAAS-2 product \cite{benas2017}, may be insufficient to fully resolve these fragmented cloud structures.

The present and upcoming generations of geostationary satellites provide continuous, high spatiotemporal resolution observations, significantly advancing our ability to study rapidly evolving dynamic systems in the Earth’s atmosphere. Historically, tracking approaches using geostationary satellite data have focused on mesoscale convective systems. Techniques for tracking clouds between successive observations range from manual methods \cite{senf_2015, senf_2017, Karagiannidis_2016, SenfDeneke2017} to automated approaches such as spatial correlation \cite{Carvalho_2001, Scofield2004, endlich_1981} and area-overlapping methods \cite{Williams_1987, Mathon_2001, vila_2008}. In some cases, a combination of correlation and overlapping techniques has been applied to improve accuracy \cite{Schroder2009}. Furthermore, fully automated tracking methods have been developed, primarily focusing on deep convective clouds to analyze mesoscale clusters \cite{Fiolleau2013} or to establish a generalized framework for diverse Earth system datasets, enhancing both adaptability and computational efficiency \cite{heikenfeld_2019}.

Individual clouds in a shallow cumulus clouds are not randomly distributed but organized into clustered patterns (e.g.~\cite{JanssensArellanoScheffer2021}). Their broken structure leads to distinct patterns of cloud shadows and illuminations \cite{Gristey2020, Lohmann2016}. \cite{Seelig_2021} uses the  Spinning Enhanced Visible and InfraRed Imager \cite{Schmit_2005} on board the European geostationary Meteosat Second Generation (MSG) satellites and employs particle image velocimetry to track these clouds over the ocean but does not account for low-level clouds' splitting and merging behaviors. This becomes an important phenomenon when the cloud grows, and merges with the neighboring clouds or a complex cloud splits up into smaller clouds. Similarly, \cite{kowch_2022} uses the Advanced Himawari Imager (AHI) rapid scan onboard HIMAWARI-8 \cite{Bessho2016} and applies the Kalman filter as a motion prediction model to estimate the locations of cloud objects across successive time frames. However, their approach assumes a constant velocity field throughout the tracking process. 
Complementing observational studies, extensive research has been conducted to track simulated shallow cumulus clouds in large-eddy simulations \cite{Zhao2005, heus_2009, heus_seifert_2013}, providing valuable insights into cloud lifecycle and dynamics.

%% file: sec-related-work.tex
\section{Related Work: Topology-Based Feature Tracking}
\label{sec:related-work}

Topology-based feature tracking for time-varying scalar fields is a two-step process: first, topological features are extracted at each time step; and second, these features are matched between adjacent time steps by solving a correspondence problem (or assignment problem in cloud science). 
Various topological descriptors---such as merge trees and persistence diagrams---have been used for feature tracking; see~\cite[Section 7.1]{YanMasoodSridharamurthy2021} for a review.
Soler et al.~\cite{SolerPlainchaultConche2018, SolerPetitfrereDarche2019} used persistence diagrams to perform topology tracking. They extracted points in a persistence diagram that encode homological features at each time step, and relied on lifted Wasserstein~\cite{SolerPlainchaultConche2018} or Wasserstein~\cite{SolerPetitfrereDarche2019} matching between persistence diagrams at adjacent time steps to establish correspondences. 
The main idea behind merge-tree-based tracking is tracking nodes of merge trees that correspond to critical points of the underlying scalar fields, and relying on matching between merge trees to establish correspondences.
Pont et al.~\cite{PontVidalDelon2021} extended the work on edit distance~\cite{SridharamurthyMasoodKamakshidasan2020} and introduced a new Wasserstein metric between merge trees to support feature tracking. 
Yan et al.~\cite{YanMasoodRasheed2022} used the labeled interleaving distance between merge trees to support geometric-aware feature tracking. 
 
Most recently, Li et al.~\cite{LiYanYan2023} introduced a probabilistic framework for tracking topological features using merge trees and optimal transport. 
In particular, they represented a merge tree as a measure network—a network associated with a probability distribution—and introduced a distance metric for comparing merge trees using partial optimal transport. This distance offers flexibility in capturing both intrinsic and extrinsic information within the comparative measures of merge trees.

A traditional method for establishing correspondence between features is to calculate the overlap between regions surrounding the features. Lukasczyk et al.~\cite{LukasczykWeberMaciejewski2017, lukasczyk2019dynamic} matched superlevel set components by measuring the overlap between their corresponding regions. 
Similarly, Saikia et al.~\cite{saikia2017global, saikia2017fast} performed topological feature tracking using merge trees. Their approach assesses the similarity of subregions segmented by merge trees at adjacent time steps, based on the overlap size between two regions and the similarity between histograms of scalar values within each region.

%% file: sec-background.tex
\section{Technical Background}
\label{sec:background}


\subsection{Merge Tree}
\label{sec:background-merge-tree}

\begin{figure}[!ht]
\centering
\vspace{-3mm}
\includegraphics[width=0.80\columnwidth]{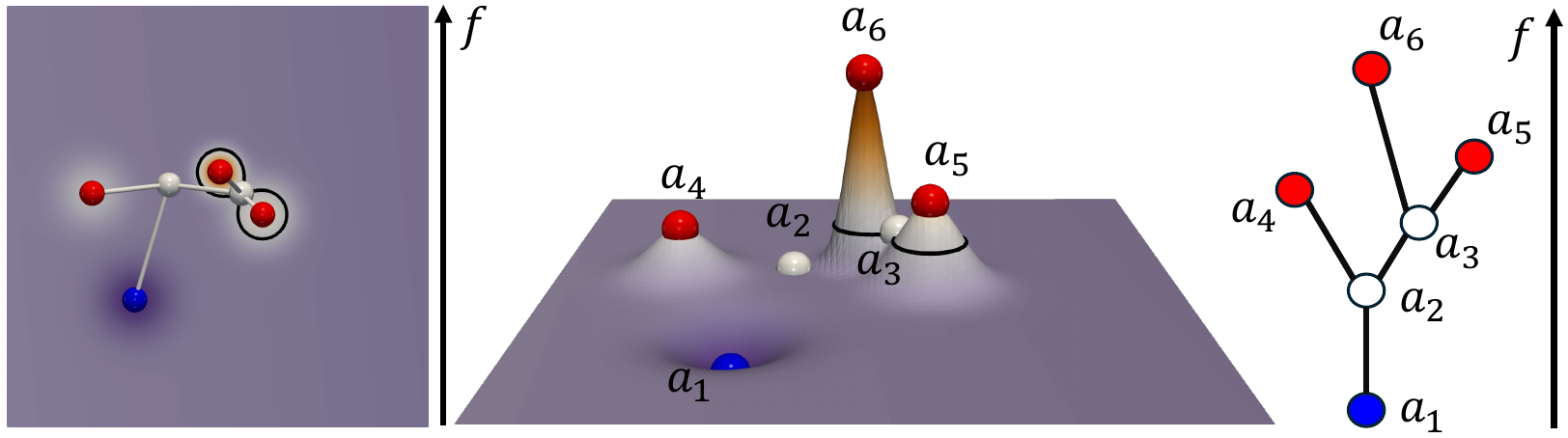}
\vspace{-3mm}
\caption{Left: a 2D visualization of a scalar field $f$ with an embedded merge tree of $-f$. Middle: a 3D visualization of the graph of $f$. Right: an abstract visualization of the merge tree of $-f$. 
Local maxima are in red, saddles are in white, and the global minimum is in blue. The black contour passing through the saddle $a_3$ encloses two peak areas of the local maxima $a_6$ and $a_5$, respectively.}
\label{fig:split-tree}
\vspace{-3mm}
\end{figure}

Let $f: \M \rightarrow \R$ be a scalar field defined on a 2D domain $\M \subset \R^2$. 
A merge tree captures the connectivity among sublevel sets of $f$. We consider two points $x, y \in {\M}$ to be  equivalent, denoted $x \sim y$, if $f(x)=f(y)=a$ and they belong to the same connected component of the sublevel set $f^{-1}(-\infty, a]$. The merge tree is a quotient space $T(\M, f)=\M/\sim$. 
For topology-based cloud tracking, $f$ corresponds to the cloud optical depth (COD) field in a satellite image with a particular timestamp; it quantifies how much a ray of light is attenuated as it travels through a cloud. A higher optical depth indicates greater extinction of light within the cloud. Since we are interested in high-value areas of $f$, we work with the merge tree of $-f$, as shown in~\cref{fig:split-tree}.  

As illustrated in~\cref{fig:split-tree}, we construct a merge tree as follows. 
We sweep the graph of $f$ (middle) with a hyperplane at the function value $a$ starting from the maximal value of $f$.
As $a$ decreases, we initiate a new branch of the merge tree each time we encounter a local maximum (e.g.,~at $a_6$, $a_5$, and $a_4$). 
Such a branch grows longer as $a$ decreases, eventually merging with another branch at a saddle point. 
For instance, at the saddle $a_3$, the branch starting at $a_6$ merges with the branch starting at $a_5$.   
Given a simply connected domain, all branches eventually merge into a single connected component at the global minimum $a_1$. Leaves, internal nodes, and the root of the tree correspond to the local maxima, saddles, and the global minimum of $f$, respectively. With a slight abuse of notation, the merge tree $T=(V,E)$ is a rooted tree whose node set $V$ is equipped with the scalar function $f$.

We define a mapping $\phi: \M \rightarrow T$ between points in the domain and the merge tree. The inverse image of an edge $e \in E$ under this mapping $\phi^{-1}(e)$ is called a \emph{topological zone}. For example, the two peak areas enclosed by the black contour in~\cref{fig:split-tree} are the topological zones of the edges $a_3a_6$ and $a_3a_5$ in $E$, respectively. 
The area of a topological zone can serve as an important measure for an edge in the merge tree~\cite{LiCarrRubel2025}, as it reflects the size of the peak in the domain. We may simplify branches with small topological zones during computation by employing this importance measure.

\subsection{Feature Tracking with Optimal Transport}
\label{sec:background-tracking}

Li et al.~\cite{LiYanYan2023} introduced topology-based feature tracking based on partial optimal transport. Our framework utilizes and significantly extends the work of Li et al.~\cite{LiYanYan2023}, making it suitable for cloud tracking. The key idea in~\cite{LiYanYan2023} is the introduction of partial Fused Gromov-Wasserstein (pFGW) distance between merge trees. The pFGW distance generates a probabilistic matching between nodes in a pair of merge trees; such a matching serves as the starting point for deriving trajectories of the cloud systems in the downstream analysis. 

\para{Optimal transport in a nutshell.} 
To illustrate optimal transport, assume there are a number of factories with specific production capacities and a number of warehouses with prescribed storage capacities. Optimal transport aims to find the most efficient way to transport goods from the factories to the warehouses by minimizing the transportation cost while respecting the capacity constraints. 
For partial optimal transport, we allow losing a certain amount of goods during transportation. 

\para{Measure network.}
Following~\cite{LiYanYan2023}, we model merge trees as measure networks. 
That is, a merge tree can be represented as a triple $T=(V, p, W)$, where $p: V \rightarrow [0,1]$ is a probability measure on the node set $V$ (i.e.,~$\sum_{x \in V} p(x) = 1$ for all $x \in V$), and $W: V \times V \rightarrow \R$ denotes the pairwise intrinsic relation between nodes. 

A measure network $T=(V, p, W)$ may also be equipped with node attributes from an attribute space $(A, d_A)$ that encodes extrinsic information. An example of a node attribute is the geometric location of its corresponding critical point in the domain. 
In this context, the attribute distance $d_A$ is the Euclidean distance between critical points in the domain.  
We will discuss our choices for $p$, $W$, and $(A, d_A)$ in the context of cloud tracking in~\cref{sec:method-pfgw}.

\para{Partial Fused Gromov-Wasserstein distance.}
The pFGW distance introduced by Li et al.~\cite{LiYanYan2023} is based on the theory of partial optimal transport~\cite{Villani2003, ChapelAlayaGasso2020}. 
Given two measure networks $T_1=(V_1, p_1, W_1)$ and $T_2=(V_2, p_2, W_2)$ equipped with node attributes, let $n_1=|V_1|$ and $n_2=|V_2|$ be the number of nodes.
A coupling $C \in \R^{n_1\times n_2}$ is a nonnegative matrix that encodes a joint probability measure between $p_1$ and $p_2$, with row and column marginals equal to $p_1$ and $p_2$, respectively.

Formally, the set of all couplings between $T_1$ and $T_2$ is
\begin{align}
\label{eq:couplings}
\Ccal = \Ccal(p_1, p_2) = \{C \in \R_{+}^{n_1 \times n_2} \mid C\onevec{n_{2}} = p_{1}, C^{\top}\onevec{n_{1}} = p_{2}\}, 
\end{align}
where $\onevec{n}=(1,1,...,1)^{\top} \in {\R}^n$.
Following \eqref{eq:couplings}, optimal transport requires a coupling (matching) to preserve all measures $p_1$ and $p_2$. 

On the other hand, partial optimal transport~\cite{ChapelAlayaGasso2020} allows partial coupling, thus partial matching between two measure networks. It relaxes the requirement for the coupling to sum to a number $m \leq 1$ (i.e.,~$m \in [0, 1]$). The set of the \emph{relaxed couplings} is
\begin{align}
\label{eq:relaxed-couplings}
&\Ccal_m = \Ccal_m(p_1, p_2) \nonumber \\
&= \{C \in \R_{+}^{n_1 \times n_2} \mid C\onevec{n_{2}} \leq p_{1}, C^{\top}\onevec{n_{1}} \leq p_{2}, \mathbf{1}_{n_{1}}^T C \onevec{n_{2}} = m\}.
\end{align}
The pFGW distance is defined on the set of  relaxed couplings $\Ccal_m$. Given a pair of merge trees modeled as measure networks $T_1$ and $T_2$, the pFGW distance is defined as 
\begin{align}
& d_q(T_{1},T_{2}) = \nonumber\\
& \min_{C\in\Ccal_m}  \sum_{ i,j,k,l} [(1-\alpha) d_A(a_i,b_j)^q 
+ \alpha |W_1(i,k) - W_{2}(j,l))|^q] C_{i,j}C_{k,l}.
\label{eq:pfgw}
\end{align}
Here, $d_A(a_i, b_j)$ is the node attribute distance between $a_i \in V_1$ and $b_j \in V_2$. 
$|W_1(i,k) - W_{2}(j,l))|$ describes the \emph{structural distortion} when we 
match pairs of nodes $(a_i, a_k) \in T_1$ with $(b_j, b_l) \in T_2$.
The pFGW distance incorporates a parameter $\alpha$ to balance the weights between these two components. 
In the context of matching a pair of merge trees, the pFGW distance provides flexibility in preserving both the node properties (such as critical point locations in the domain) and the merge tree structure in the optimal coupling. It also allows the appearance and disappearance of new features.

In the ``factory-warehouse'' scenario, we are in the setting of optimal transport when $m=1$ in~\eqref{eq:pfgw}. 
$p_1$ and $p_2$ prescribe the capacities of factories and warehouses, respectively, and the coupling $C$ describes a transportation plan that respects these capacity constraints. The transportation cost is described by the attribute distances (e.g.,~Euclidean distances) between factories and warehouses as well as the structural relations among them (e.g.,~factories owned by a given company should transport goods to warehouses owned by the same company).  
Solving an optimization problem of~\eqref{eq:pfgw} means finding the transportation plan with the lowest cost.  
On the other hand, we are in the setting of partial optimal transport when $0<m<1$ in~\eqref{eq:pfgw}, meaning that we allow $1-m$ percent of goods to be lost/ignored during transportation.

%% file: sec-method.tex
\section{Method}
\label{sec:method}

Cloud tracking faces three major challenges. First, clouds observed in satellite images are complex, time-varying phenomena involving numerous events, as cloud systems appear, disappear, merge, and split. Second, there is no consensus among domain scientists on the definition of cloud objects and cloud systems. Third, there are no ground-truth cloud tracking results available for satellite images supporting any form of supervised learning. 

In this section, we describe our novel framework of topology-driven cloud tracking. Working closely with domain scientists, we first introduce the definition and detection of cloud objects (\cref{sec:method-detect-cloud-objects}). 
We then describe our strategy of using critical points as anchor points for cloud objects  (\cref{sec:method-anchor-point}). 
Subsequently, we compute a matching between the anchor points using partial optimal transport (\cref{sec:method-pfgw}). 
We then generate trajectories for cloud systems formed by (possibly) multiple cloud objects (\cref{sec:method-cloud-system}). 
Implementation details are in the supplement. 

\subsection{Detecting and Simplifying Cloud Objects}
\label{sec:method-detect-cloud-objects}

In a cloud optical depth (COD) field $f$ from a satellite image, regions with high function values usually indicate thicker clouds.


\para{Cloud object detection.} 
We use a thresholding strategy for the detection of cloud objects. We define each connected component of a superlevel set of $f$ at a chosen threshold $a$ (i.e.,~$f^{-1}[a, \infty)$) as a \emph{cloud object}. Currently, there is no consensus on the threshold value $a$ to detect low-level clouds. This results in a lack of widely accepted ground-truth data for cloud detection and tracking. Following established practices~\cite{heikenfeld_2019}, we test a range of thresholds from $0.5$ to $5.0$ at a gap of $0.5$ to analyze the impact of the threshold on (a) the number of cloud objects, and (b) their size distributions. We choose a threshold $a = 2.0$ to avoid under- and over-segmentation; see supplement (parameter sensitivity analysis) for details.   

\para{Cloud object simplification.} 
We want to separate features from noise in our real-world cloud data. 
Stratocumulus clouds often cover large, continuous areas that can stretch hundreds of kilometers; therefore, we may consider removing smaller cloud objects that are deemed insignificant from stratocumulus clouds. 
On the other hand, shallow cumulus clouds are characterized by their small size (covering hundreds of meters across), relatively flat bases, and puffy tops; they could also grow into deeper convective systems depending on the available convective mass flux, as seen in their COD values~\cite{vial_2022}. To handle these differences, we use different approaches to simplify the identification of cloud objects for stratocumulus and shallow cumulus, respectively. 

\begin{figure}[!ht]
\vspace{-3mm}
\centering
\includegraphics[width=0.70\columnwidth]{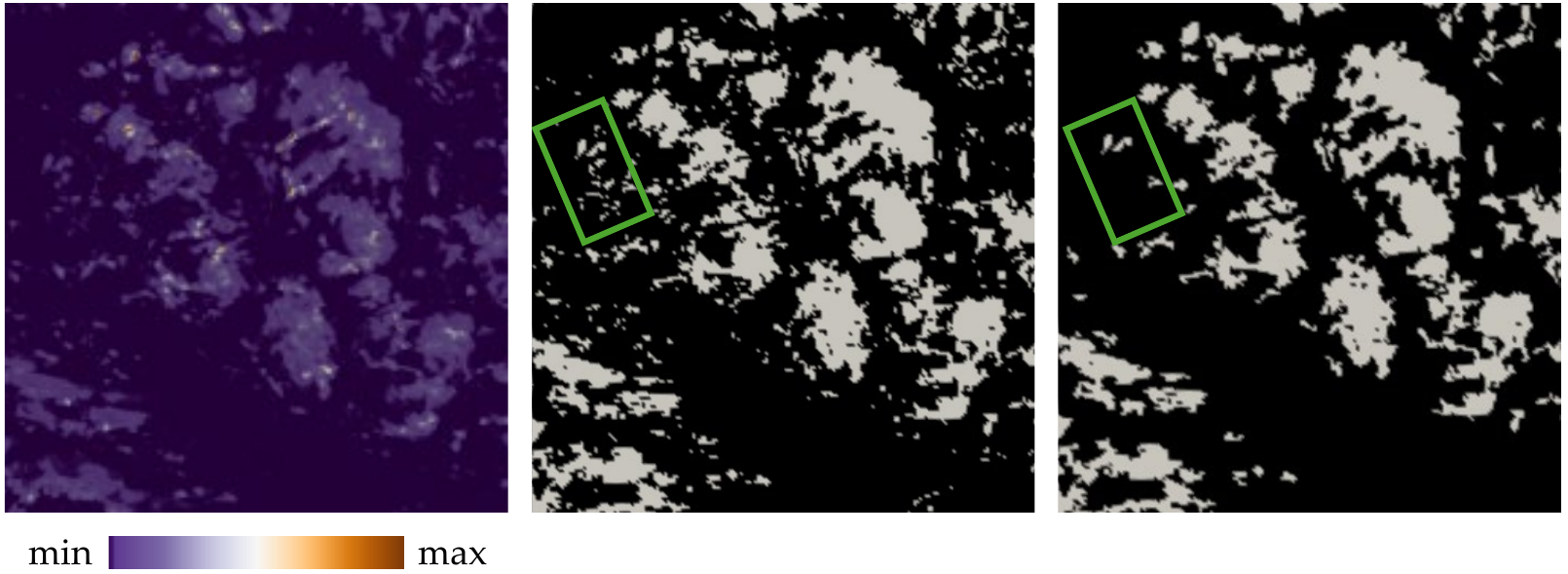}
\vspace{-3mm}
\caption{Apply area-based simplification to stratocumulus clouds. From left to right: the COD field, the cloud area map, and the simplified map by excluding cloud objects smaller than 10 pixels.}
\label{fig:cloud-simplification}
\vspace{-3mm}
\end{figure}

For stratocumulus (typically over the ocean), we ignore small cloud objects based on their coverage area (size); see~\cref{fig:cloud-simplification}. 
We use the statistics of cloud area size to determine the area-based simplification level: by ignoring cloud objects smaller than $10$ pixels, we can get rid of more than $70\%$ of cloud objects from the field and still cover more than $97\%$ of the total cloud area; see supplement for details. \cref{fig:cloud-simplification} gives an example of area-based simplification. The cloud area maps show cloud objects in gray and the background in black. 
In the simplified cloud area map, cloud objects smaller than $10$ pixels are ignored after simplification (cf.~the green boxes). 

\noindent For Shallow cumulus (typically over land), instead of filtering by size, we apply a higher COD threshold to remove regions with low values, focusing only on the prominent clouds; see supplement for details. 

\subsection{Attaching Anchor Points to Cloud Objects}
\label{sec:method-anchor-point}
We need to associate cloud objects with topological features to perform topology-driven tracking. 
To that end, we use nodes of merge trees that correspond to the critical points of the COD field $f$ as anchor points for cloud objects. 

\begin{figure}[!ht]
\centering
\includegraphics[width=0.75\columnwidth]{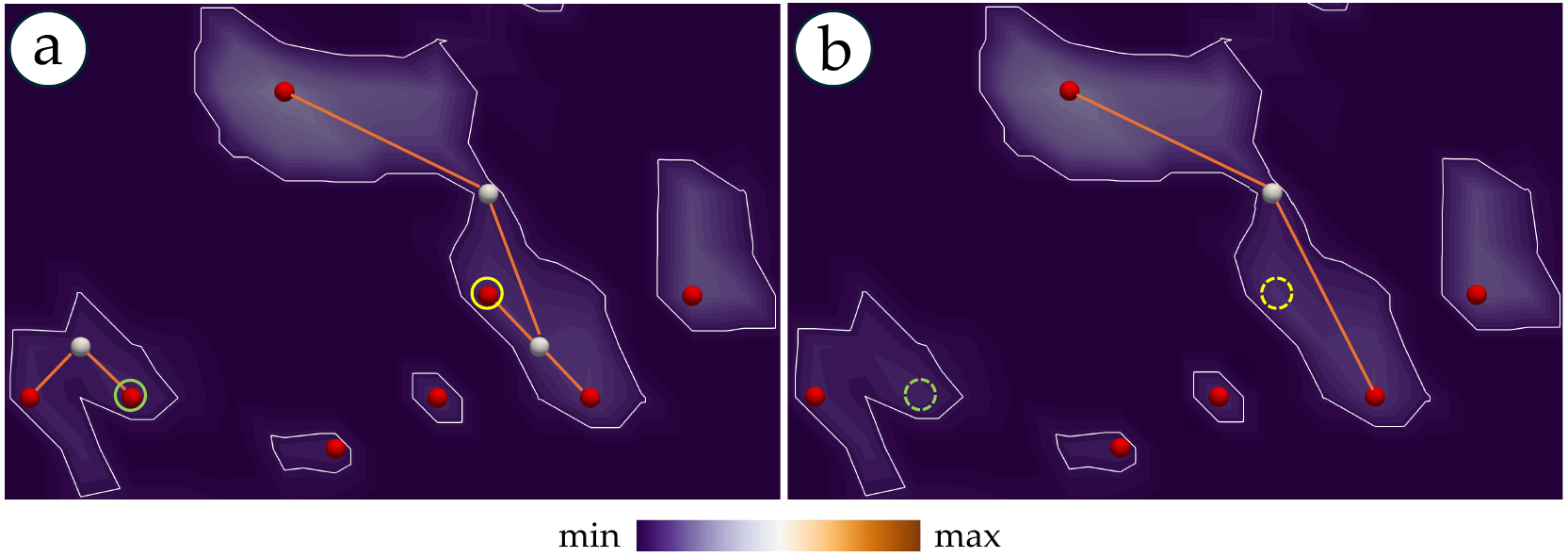}
\vspace{-3mm}
\caption{(a) A set of cloud objects enclosed by white contours; each contains a subtree of the global merge tree. Local maxima (a.k.a.,~anchor points) are in red, and saddles are in white. (b) Simplifying subtrees by removing the highlighted anchor points (inside green or yellow circles) and their parent saddles.}
\label{fig:intra-cloud-simplification}
\vspace{-6mm}
\end{figure}

To attach anchor points, we could associate a subtree of the merge tree to each cloud object. 
For example,~\cref{fig:intra-cloud-simplification}(a)  shows five cloud objects enclosed by the white contours. The tree structure within each cloud object is a subtree of the global merge tree. 
We use the local maxima in this subtree as the anchor points of the cloud object. 
These anchor points' trajectories are subsequently used to derive the trajectory of cloud objects. 
In practice, we may reduce the number of anchor points for computational efficiency; see~\cref{fig:intra-cloud-simplification}(b) for an example and technical details in the supplement. 

\subsection{Anchor Point Tracking with Partial Optimal Transport}
\label{sec:method-pfgw}

We adapt the work of Li et al.~\cite{LiYanYan2023} to track anchor points with partial optimal transport. Building on prior knowledge of the characteristics of the COD field, we enhance this work by incorporating a tailored probability distribution for critical points. 

\para{Model merge trees as measure networks.}
We first model a merge tree $T$ of the COD field $f$ as an attributed measure network $T=(V, p, W)$ with node attributes $(A, d_A)$. We modify the framework of~\cite{LiYanYan2023} to focus on tracking local maxima of a merge tree, which act as anchor points for cloud systems. 

For each local maximum $x\in V$, we set its probability as $p(x) = 1/n$, where $n$ is the number of local maxima in $T$. For saddles and the global minimum, we assign a probability of $0$ due to the high complexity and uncertainty of the COD field~\cite{Platnick2017}, which causes their locations to be highly unstable. This instability makes it challenging to find suitable matchings for these nodes. 
Furthermore, disregarding the probability of saddles and the global minimum enables us to preserve more anchor points without compromising computational efficiency, ultimately enhancing the robustness of tracking.

We use the pairwise node relation matrix $W$ to encode the tree distance. 
Recall that each node $v \in V$ is equipped with a function value $f(v)$. 
The tree distance between two adjacent nodes in $T$ (\ie, the edge weight) is defined as $W(a, b) = |f(a)-f(b)|$, whereas the tree distance between two nonadjacent nodes is the shortest path distance between them. 
Previous works~\cite{LiPalandeYan2023, LiYanYan2023} have shown that the tree distance can encode the scalar field topology via the merge tree structures. Specifically, in the context of cloud tracking, the tree distance reflects the locality among the anchor points: anchor points attached to the same cloud object belong to the same subtree. 
Our framework captures anchor point locality inherently, regardless of whether saddles or the global minimum are preserved in the optimal transport. 

We encode the locations of critical points as the node attributes. 
Given a pair of merge trees $T_1=(V_1, p_1, W_1)$ and $T_2=(V_2, p_2, W_2)$, the node attribute for $a_i \in V_1$ is $(x_i, y_i)$, in which $x_i$ and $y_i$ denote the coordinates of the critical point. Similarly, the node attribute for $b_j \in V_2$ is $(x_j, y_j)$. The attribute distance $d_A$ between $a_i$ and $b_j$ is 
\begin{equation}
d_A(a_i, b_j) = d_E\left((x_i,x_j), (y_i, y_j)\right)  
\label{eq:attribute}
\end{equation}
$d_E$ in~\cref{eq:attribute} represents the Euclidean distance and our framework  prevents the matching of anchor points that are far apart. 

\para{Matching critical points with partial optimal transport.}
By computing the pFGW distance between a pair of merge trees $T_1$ and $T_2$ following \eqref{eq:pfgw}, we obtain an optimal coupling $C$ between their nodes.
We interpret the coupling as a probabilistic matching between critical points from adjacent time steps. 
The sub-matrix of the coupling matrix, consisting only of rows and columns corresponding to local maxima, represents the matching between anchor points. 

\begin{figure}[!ht]
\centering
\vspace{-3mm}
\includegraphics[width=0.8\columnwidth]{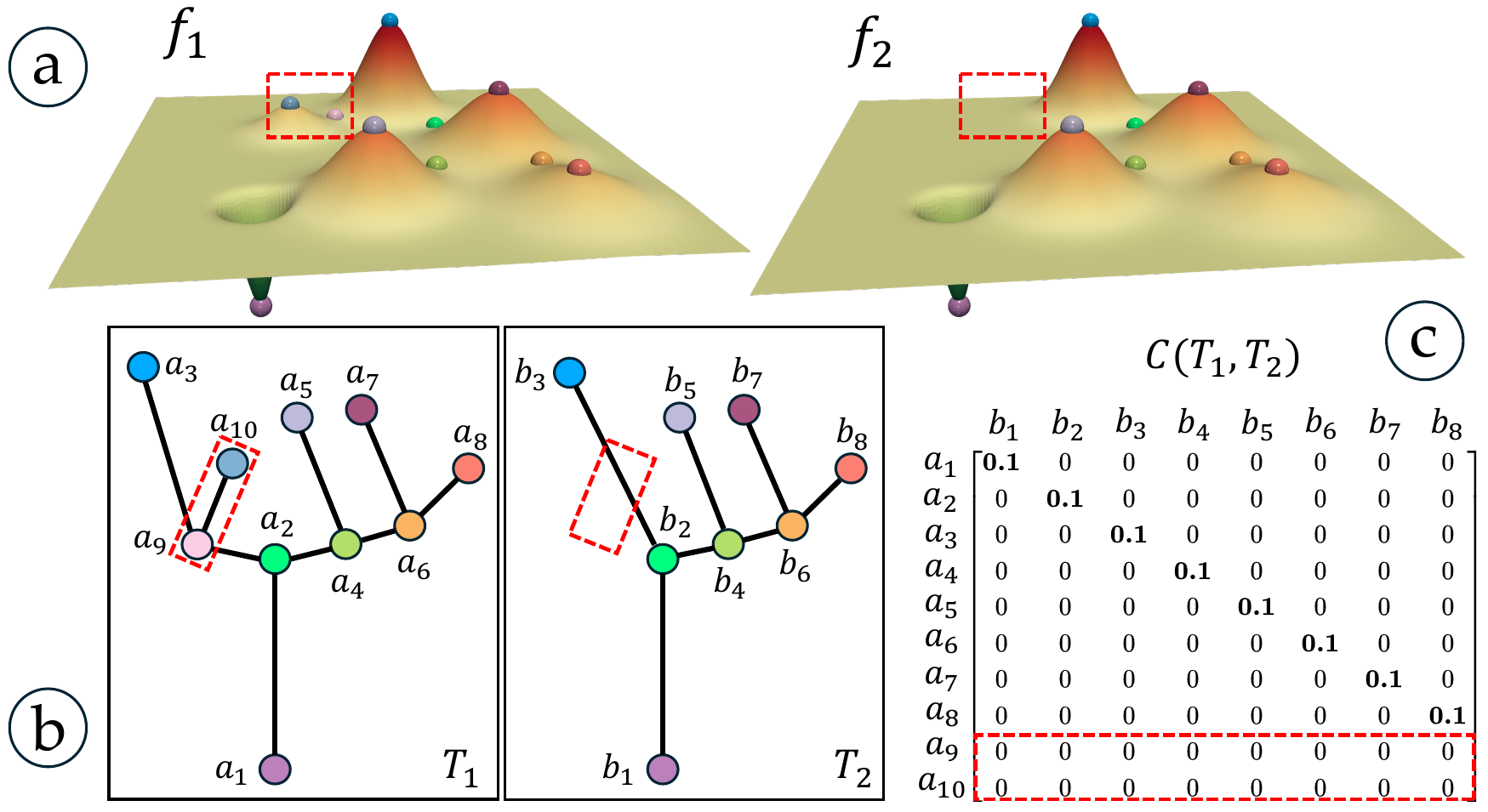}
\vspace{-3mm}
\caption{Partial optimal transport. We match the merge trees of  $-f_1$ and $-f_2$ (b) using using pFGW with $m=0.8$, producing a coupling matrix $C$ (c). Matched nodes in the two trees share the same color (a-b).}
\label{fig:pfgw-example}
\vspace{-3mm}
\end{figure}

We provide a simple example in~\cref{fig:pfgw-example}.  
Let $f_1$ and $f_2$ in (a) be the scalar fields of adjacent time steps. 
$T_1$ and $T_2$ in (b) are the merge trees of $-f_1$ and $-f_2$, respectively. 
Let $p_1$ and $p_2$ be uniform measures on all the nodes, including local maxima, saddles, and the global minimum. 
Based on structural similarity between $T_1$ and $T_2$, it is natural to match nodes $a_i \in T_1$ with $b_i \in T_2$ for $i \in [1,8]$. 
On the other hand, nodes $a_9$ and $a_{10}$ in $T_1$ are missing from $T_2$ (c.f.,~the red boxes). 
Ideally, we want to ignore $a_9$ and $a_{10}$ during the matching process. Based on these intuitions, setting $m=0.8$ in the pFGW distance produces the desired transportation plan captured by $C = C(T_1, T_2)$ in (c).  

Each entry $C_{i, j}$ in $C$ denotes the probability of matching $a_i \in T_1$ with $b_j \in T_2$.
We see that $C_{i,i} = 0.1$, which aligns well with our expectations. Meanwhile, $C_{9,j} = C_{10,j} = 0$ for $j \in [1,8]$, indicating that $a_9, a_{10} \in T_1$ are not matched to any nodes in $T_2$. 
A sub-matrix of \( C \), with rows corresponding to \( a_3 \), \( a_5 \), \( a_7 \), \( a_8 \), \( a_{10} \) and columns corresponding to \( b_3 \), \( b_5 \), \( b_7 \), \( b_8 \), represents the probabilistic matching between anchor points. 
While this example shows a one-to-one node matching, our framework generally allows multiple nonzero entries in a row or column,
which distinguishes our framework from other topology-based frameworks that produce one-to-one critical point matchings. 

\subsection{Computing Trajectories for Cloud Systems}
\label{sec:method-cloud-system}

\para{Obtaining cloud systems.}
We first merge cloud objects into cloud systems.
A cloud system may consist of multiple cloud objects that are geometrically close. 
Eytan et al.~\cite{EytanKorenAltaratz2020} suggested that most of the radiative effect of a cloud is confined within \textasciitilde $4$km around the cloud.  Therefore, we merge cloud objects within $4$km away from each other as a cloud system and identify cloud objects farther than $4$km as different cloud systems. 
The set of anchor points for each cloud system is the collection of anchor points for all cloud objects within the system.

We do not define cloud systems when identifying cloud objects (\cref{sec:method-detect-cloud-objects}). This is because a subset of the global merge tree within a cloud system may be a forest, making intra-cloud anchor point simplification slightly more complex. 

\para{Tracking cloud systems.}
As described in~\cref{sec:method-anchor-point}, each cloud system contains one or more local maxima as its anchor point. 
We can use the matching between the anchor points (see~\cref{sec:method-pfgw}) to compute the trajectory for cloud systems.

We introduce a matching score between cloud systems at adjacent time steps. 
We denote the set of anchor points for a cloud system $X$ as $P_X$. The matching probability from the optimal coupling between an anchor point $v_1$ at time step $t$ and $v_2$ at time step $(t+1)$ is $C_{t}(v_1, v_2)$. 
Then, for the cloud systems $X$ (at time step $t$) and $Y$ (at time step $(t+1)$), the matching score between them is \begin{equation}
S_t(X, Y) = \sum_{x\in P_X, y\in P_Y}C_{t}(x, y). 
\label{eq:matching-score}
\end{equation} 
Informally, this score is the probability of mass transportation from $X$ to $Y$. The higher this score is, the more likely the two are matched.

\begin{figure}[!ht]
\centering
\vspace{-3mm}
\includegraphics[width=0.8\columnwidth]{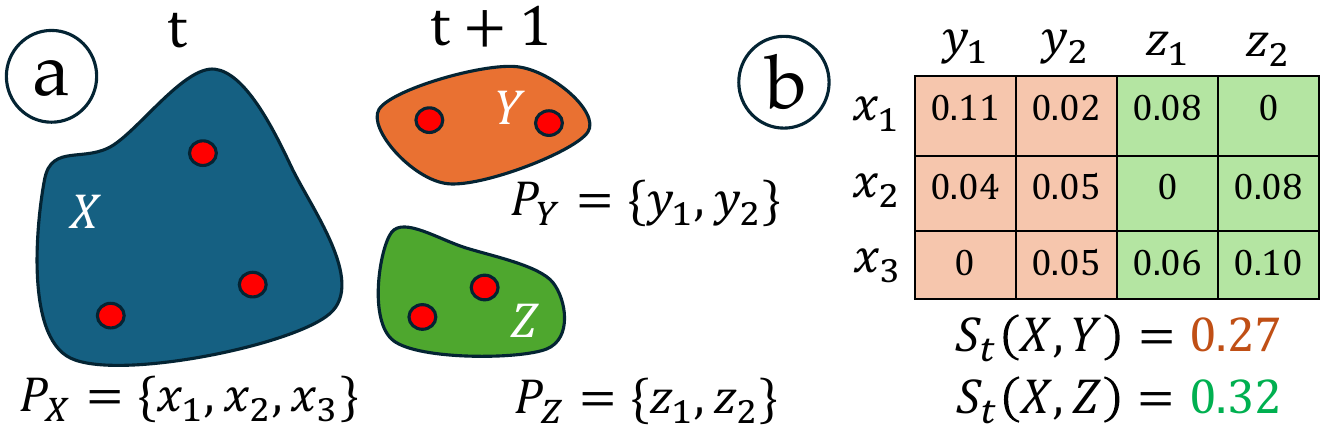}
\vspace{-3mm}
\caption{Cloud system matching score. 
(a) Cloud system $X$ at time step $t$ and cloud systems $Y$ and $Z$ at time step $(t+1)$, along with their set of anchor points.
(b) Selected rows and columns of the coupling matrix $C$.}
\label{fig:cloud-system-matching-score}
\vspace{-3mm}
\end{figure}

\cref{fig:cloud-system-matching-score} gives an example of matching scores involving cloud systems $X$, $Y$, and $Z$. 
In (a), there are three anchor points for $X$ at time step $t$, and two for $Y$ and $Z$ at time step $(t+1)$, respectively. 
The matrix $C$ for the matching probability between the anchor points is shown in (b). For example, the matching probability between anchor point $x_1$ from $X$ and $y_1$ from $Y$ is $0.11$.
The matching score $S_t(X, Y)$ equals the sum of the first two (orange) columns for anchor points $y_1$ and $y_2$, which is $0.27$. 
Similarly, $S_t(X, Z)$ equals the sum of the last two (green) columns for anchor points $z_1$ and $z_2$, which is $0.32$.

With the matching scores, we can match the cloud systems via a bipartite graph matching algorithm.
Let $H_t$ denote the set of cloud systems at time step $t$. 
We use $S_t(X,*)=\sum_{h\in H_{t+1}} S_t(X, h)$ to denote the total outgoing probability for a cloud system $X \in H_t$ and $S_t(*,Y)=\sum_{h\in H_{t}} S_t(h, Y)$ the total incoming probability for $Y \in H_{t+1}$.  
A matching between the cloud system $X$ and $Y$ is \emph{valid} if $S_t(X, Y)$ is nonzero and satisfies one of the two conditions:
\begin{enumerate}[noitemsep,leftmargin=*]
    \item $Y = \mathop{\arg\max}\limits_{h\in H_{t+1}} S_t(X, h)$, and $X = \mathop{\arg\max}\limits_{h\in H_{t}} S_t(h, Y)$;
    \item $S_t(X, Y) \geq \max\{S_t(X, *), S_t(*, Y)\} \times r$.
\end{enumerate}
Condition (1) means that \( X \) and \( Y \) are mutually the best match; otherwise, condition (2) implies that the matching score must exceed a threshold proportional to the maximum cumulative score of both \( X \) and \( Y \). The proportionality factor is governed by the parameter \( r \), which controls the strictness of the matching criteria. In practice, we set $r=0.1$. We justify this parameter choice in the supplement.

Among all valid matchings, we search for a one-to-one cloud system matching strategy that prioritizes the pairing of cloud systems with larger areas. We implement this process using a greedy algorithm.
\begin{enumerate}[noitemsep,leftmargin=*]
\item 
First, we sort all cloud systems \( X \in H_t \) in descending order based on their areas, ensuring that larger cloud systems are processed first.
\item 
For each cloud system $X$ in the sorted list, we evaluate all candidate cloud systems $Y\in H_{t+1}$
  that satisfy matching conditions (1) or (2) and choose the one with the largest area to be matched to $X$.
\item 
For all remaining cloud systems that are unmatched, we mark them as terminated (for $X \in H_t$) or newly formed (for $Y \in H_{t+1}$).
\end{enumerate}
By combining all the selected matchings across time steps, we generate a set of trajectories for the cloud systems. 

\para{Merge and split events.}
Oftentimes, cloud systems merge and split as they evolve over time, offering insights to weather scientists into their evolution. 
Our framework supports computing and visualizing these events. 
Following the tracking algorithm outlined above, we have computed all valid matchings, with one labeled as the main trajectory and the others identified as secondary trajectories. 
In the example in~\cref{fig:cloud-system-matching-score}, we let the main trajectory of the cloud system $X$ go to $Z$ if the area of $Z$ is larger than $Y$.  
However, the trajectory from $X$ to $Y$ can also be considered secondary due to the high matching score between $X$ and $Y$. 
Including this secondary trajectory allows us to interpret the scenario as follows: cloud system $X$ at time step $t$ splits into two systems, $Y$ and $Z$, at time step $(t+1)$, with $Z$ following the main trajectory of the system.

\subsection{Algorithmic Highlights}
In~\cref{sec:method}, we introduce our novel framework that extends the general topology-based tracking tool using the pFGW distance~\cite{LiYanYan2023}. 
Specifically, we apply a customized probability distribution for anchor points, informed by existing knowledge of the cloud data. Additionally, we propose the concept of tracking cloud systems and introduce a matching algorithm for cloud systems based on optimal transport. 

%% file: sec-results.tex
\section{Experimental Results}
\label{sec:results}

In this section, we experiment with two datasets from geostationary satellites. The first {\DM} dataset focuses on marine stratocumulus clouds over the ocean west of Africa during August 2023, whereas the second {\DL} dataset covers shallow cumulus cloud systems over central Europe from April to September between 2018 and 2019; see supplement for details on these datasets. We review and compare against two state-of-the-art cloud tracking tools (\cref{sec:existing-tools}), with parameter justifications (\cref{sec:results-parameter}). We then perform statistical evaluations (\cref{sec:evaluation-metric}) and discuss our tracking results in~\cref{sec:result-d1,sec:result-d2}. 
Additionally, we compare our approach with two topology-based general-purpose tracking tools in~\cref{sec:result-topology-tools}. 

\subsection{Two Leading Cloud Tracking Tools}
\label{sec:existing-tools}

We report the results from two state-of-the-art open-source cloud tracking tools for comparative analysis: {\tobac}~\cite{heikenfeld_2019, SokolowskyFreemanJones2024} and {\PyFT}~\cite{FengHardinBarnes2023}. 
We refer to our tool as the {\pFGW} framework.

The tool {\tobac} takes a sequence of thresholds to identify connected components of the superlevel set of the COD field. Specifically, for a fixed threshold, it calculates the bounding box of each superlevel set component and selects a feature point from the bounding box using one of four strategies: the center, the maxima, or the barycenter weighted by either the absolute COD value or its difference from the threshold.
Then, {\tobac} uses the watershed algorithm~\cite{NAJMAN1994} to detect cloud objects. 
The watershed algorithm first identifies the local peak area for each feature point. 
A cloud object is then created at the local maxima of the identified peak area and expanded by iteratively adding the surrounding pixels with the highest COD values.
The expansion terminates when the cloud object touches another one or reaches the boundary of the superlevel set component. 
Subsequently, the trajectory of the cloud object is defined by the trajectory of the feature point. 
To match a feature point to one in the next time step, {\tobac} searches for possible candidates within a user-defined neighborhood in the domain. 
The matching strategy that minimizes the sum of the squared Euclidean distance between the feature point and its matched point produces the tracking result. 

{\PyFT} identifies cloud objects using either superlevel set components or the watershed algorithm.
Then, it computes the region overlap between cloud objects at adjacent time steps and determines the trajectory of the cloud object based on the overlapped region size. 

For cloud detection, {\tobac} employs the watershed algorithm to expand each cloud from its respective feature point. This expansion is essential as {\tobac} relies on feature points to facilitate tracking in subsequent stages. 
In particular, if one can guarantee that each superlevel set component has exactly one feature point inside, the cloud object detection result is identical to the superlevel set component. 
However, since {\tobac} selects the feature point from the component's bounding box, none of the four feature detection strategies can guarantee this outcome. 
In contrast, {\pFGW} identifies cloud objects using superlevel set components, whereas  {\PyFT} offers flexibility by supporting both superlevel sets and the watershed algorithm for cloud detection. 

During cloud tracking, {\pFGW} combines topological and geometric information of anchor points and summarizes the tracking results for all anchor points within a cloud system. In contrast, both {\tobac} and {\PyFT} rely exclusively on geometric information. {\tobac} tracks clouds using a single feature point for each cloud object; it may produce an unstable trajectory due to the instability in a feature point's location across time steps.  On the other hand, {\PyFT} tracks features based on region overlap; however, it could be challenging to handle small or fast-moving features with insufficient overlaps between adjacent time steps~\cite{LiYanYan2023}. 

Furthermore, instead of tracking cloud objects, {\pFGW} considers multiple cloud objects as a cloud system and tracks the system as a whole. For simplicity, we use \emph{cloud entity} to refer to either cloud object or cloud system for the rest of the paper. 

\subsection{Parameter Configurations}
\label{sec:results-parameter}
Computing the {\pFGW} distance requires two parameters $\alpha$ and $m$ (see~\cref{sec:background-tracking}). We set $\alpha=0.4$ for the {\DM} dataset and $\alpha=0.2$ for the {\DL} dataset. 
For a matching between adjacent time steps, we use an automatic process to find the highest value of \( m \), provided that no Euclidean distance between matched anchor points exceeds a user-defined threshold; see the supplement for a discussion about parameter choices. 
For {\tobac}, we use the barycenter of a superlevel set component as the feature point, which is reported to provide the optimal tracking results~\cite{heikenfeld_2019, SokolowskyFreemanJones2024}. 
For {\PyFT}, we choose the strategy based on superlevel set components to be consistent with our {\pFGW} approach.
For a fair comparison, we use the same threshold for cloud detection for all three methods. 
Additional parameters for the three methods and a discussion of limitations are in the supplement.


\begin{figure*}[!ht]
\centering
\includegraphics[width=1.6\columnwidth]{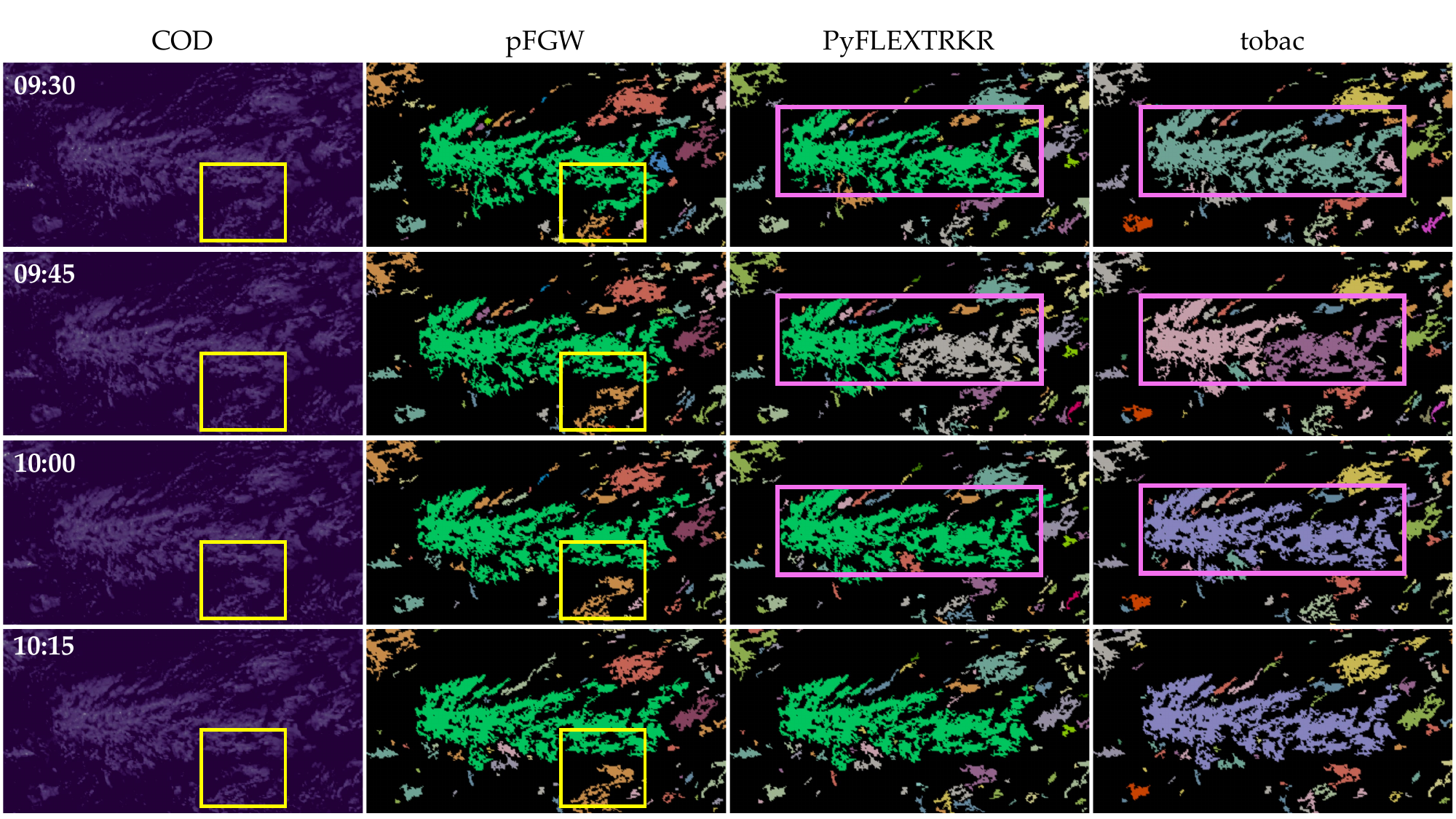}
\vspace{-3mm}
\caption{Tracking results of a region in the {\DM} dataset on Aug 1, 2023, at 9:30, 9:45, 10:00, and 10:15 UTC, respectively. Cloud entities are colored by feature correspondences. 
From left to right: visualizations of COD fields, tracking results for {\pFGW}, {\PyFT}, and {\tobac}, respectively. 
For {\pFGW}, yellow blocks from top to bottom showcase a cloud transition process from the green cloud system to the orange one. 
For {\PyFT} and {\tobac}, magenta boxes highlight suboptimal tracking results due to the transient splitting and merging of cloud objects.} 
\vspace{-3mm}
\label{fig:D1-case-study}
\end{figure*}

\subsection{Evaluation Metrics}
\label{sec:evaluation-metric}

Tracking clouds over time using satellite observations is challenging due to their dynamic nature, including their appearances, disappearances, splitting, merging, and transformation. Based on earlier studies in cloud science~\cite{Fiolleau2013, valliappa_2010}, we utilize three evaluation metrics to assess the tracking results.  

First, we study the distribution of \textbf{timespans} for cloud entities (cloud objects or cloud systems). 
The timespan is the duration of time (i.e., the number of time steps) a cloud entity travels along its trajectory. This metric indicates how consistently a tracking method monitors cloud entities. However, we do not postulate that every cloud entity should be long-lived.  

Second, we investigate the distribution of the standard deviation of a cloud property (e.g., mean COD value of a cloud entity at a given time step) along trajectories, referred to as the \textbf{SD of mean COD}. The physical properties of a cloud entity are expected to remain fairly stable over time, with no significant deviations. 
A mismatch is likely to result in an increase in the standard deviation along the trajectory. 
We consider only trajectories with a timespan longer than the median timespan and at least three time steps, as the standard deviation is highly sensitive to small sample sizes.

Third, we examine the distribution of the \textbf{linearity loss} of trajectories.  
The linearity loss of a trajectory is the root mean square error (RMSE) of the centroids of cloud entities from the line of best fit. 
Given the momentum of clouds, it is reasonable to expect short-lived trajectories to be nearly linear without abrupt jumps. 
However, longer trajectories are more likely to follow the mean flow, which can vary across time and space, resulting in curved trajectories that are not accounted for by this metric. 
For this metric, we focus on the same set of trajectories as those considered in the {\CODm} study.

\subsection{Case Study: {\DM} Dataset}
\label{sec:result-d1}

We first highlight our topology-driven tracking results in~\cref{fig:teaser} using two time steps from the {\DM} dataset on Aug 1, 2023, at 09:00 and 10:00 UTC, respectively.  
We then perform a detailed analysis of the cloud tracking results using a subregion from the same dataset also on Aug 1, 2023 in~\cref{fig:D1-case-study}. There are $28$ time steps within the day, captured from 09:00 to 15:45 UTC with $15$-minute intervals. 

\para{Cloud detection and tracking.}
We first compare the cloud detection results across the three cloud tracking tools: {\pFGW}, {\PyFT}, and {\tobac}.
All three methods successfully identify cloud entities from the COD fields, with small discrepancies due to the minor differences between watershed-based (used by {\tobac}) and superlevel-set-based (used by {\PyFT} and to some extent {\pFGW}) strategies.  

Meanwhile, we highlight the differences between tracking \emph{cloud objects} versus \emph{cloud systems} in~\cref{fig:D1-case-study}. 
For {\PyFT} (3rd column), at 09:45 UTC, the central green object (magenta box) splits into two distinct objects (green and gray). At 10:00 UTC, these two objects merge back together, and the trajectory of the newborn gray object terminates. 
For {\tobac} (4th column), we observe similar cloud splitting and merging events at 09:45 and 10:00 UTC, respectively; however, {\tobac} considers the objects (magenta boxes) at 9:45 and 10:00 UTC to be new entities, giving rise to three new trajectories.  
However, the cloud-splitting event at 09:45 UTC is not obvious in the COD field. 
In contrast, at 09:45 UTC, {\pFGW} does not split the same green cloud system in the center, as our tracking method aggregates nearby cloud objects into a single cloud system. 

Previous studies have shown significant COD uncertainties beyond the $5 - 50$ value range~\cite{Platnick2017}. With a cloud detection threshold of 2.0, these COD uncertainties can blur the cloud object boundaries, causing such transient splitting and merging events. 
An advantage of tracking cloud systems (instead of cloud objects) with {\pFGW} is that it avoids generating many short-lived trajectories for these transient events.

As shown in~\cref{fig:D1-case-study}, we gain additional insights by tracking cloud systems instead of cloud objects using {\pFGW}. The yellow boxes in the 1st and 2nd columns highlight a cloud transition process, where a part of the central green system splits and merges into the bottom orange system. 
In contrast, {\PyFT} and {\tobac} track all cloud objects individually in this region, thus it is harder to infer the change in proximity between clouds.

\para{Statistical evaluation.}
We statistically evaluate the {\DM} dataset from Aug 1 to Aug 8, 2023. The observed time period for each day is from 09:00 to 15:45 UTC with a $ 15$-minute interval. 
We calculate the tracking results for each day separately and then aggregate the statistics from all eight days to evaluate the overall performance for each method.
For {\pFGW}, we include statistics involving tracking cloud systems (\textit{pFGW-system}) and tracking cloud objects (\textit{pFGW-object}). 

\begin{figure}[!ht]
\vspace{-1mm}
\centering
\includegraphics[width=0.75\columnwidth]{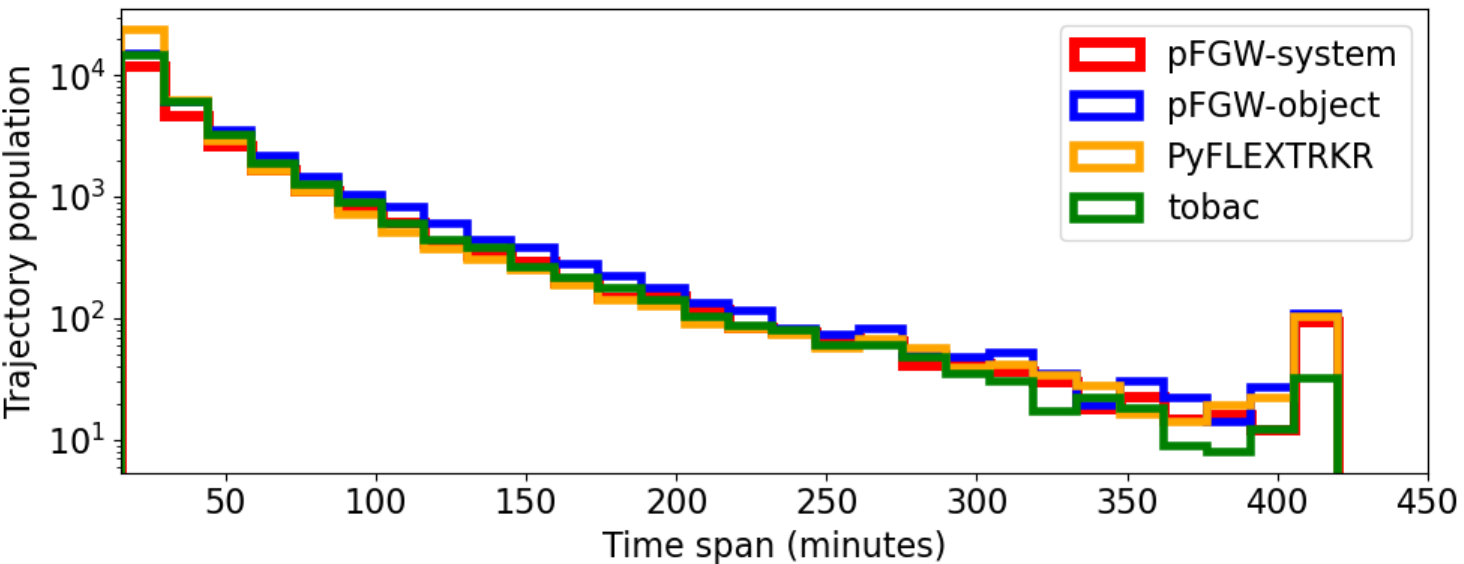}
\vspace{-2mm}
\caption{{\DM} dataset: distribution of trajectory timespans in log-scale for {\pFGW} tracking cloud systems (red) and objects (blue), {\PyFT} (orange), and {\tobac} (green); data aggregated over eight days (Aug 1-8, 2023). 
}
\label{fig:D1-timespan}
\vspace{-3mm}
\end{figure}

\cref{fig:D1-timespan} shows the distributions of trajectory timespan for all three methods. These distributions are comparable across all three methods. The distributions  of {\pFGW} for tracking cloud objects and tracking cloud systems are also similar. 
Specifically, {\PyFT} generates more short-lived trajectories (primarily those lasting less than 15 minutes); {\tobac} generates fewer longer-lived trajectories compared to the other two methods.
In comparison, {\pFGW} is able to preserve long-term trajectories and reduce the amount of short-lived trajectories.

\begin{figure}[!ht]
\vspace{-1mm}
\centering
\includegraphics[width=0.95\columnwidth]{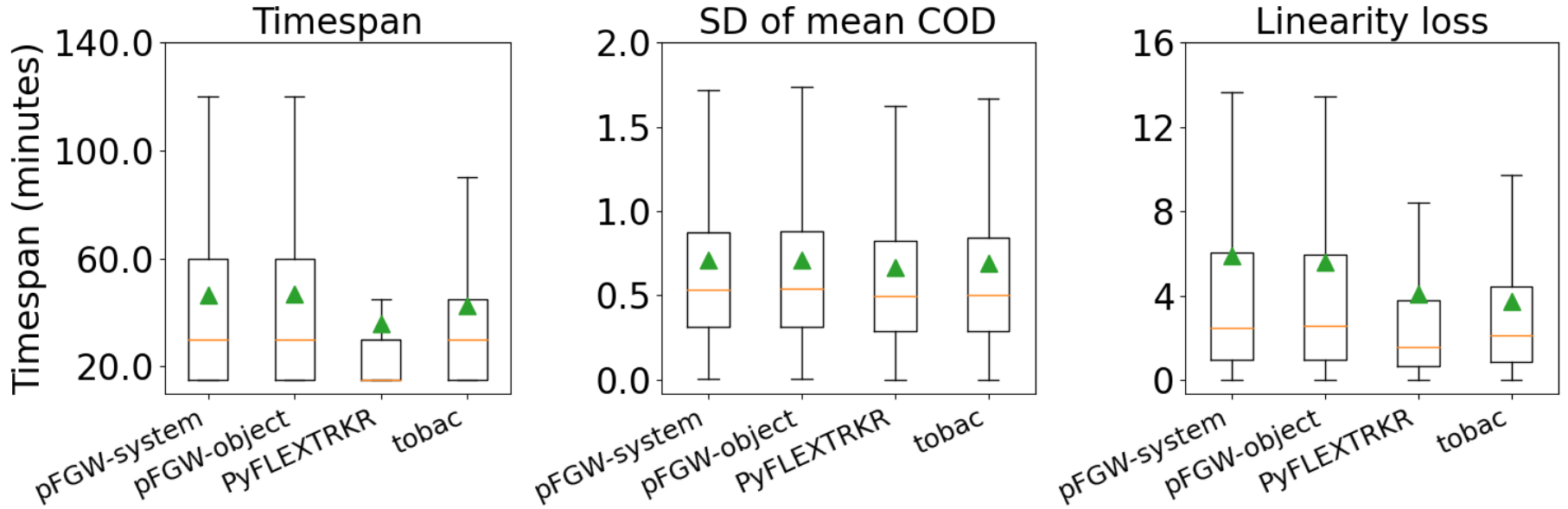}
\vspace{-3mm}
\caption{{\DM} dataset: box plots showing the median (orange line), mean (green triangle), and interquartile range (box boundary) of the distribution for three evaluation metrics.}
\label{fig:D1-statistics}
\vspace{-3mm}
\end{figure}

\cref{fig:D1-statistics} presents the overall statistics for comparison and displays the \emph{trajectory timespan} distribution in the form of a box plot (1st column). {\PyFT} generates more short-lived trajectories compared to the other two methods, with the median trajectory timespan being just 15 minutes (one timestep). 
In contrast, {\pFGW} and {\tobac} exhibit a higher median value of 30 minutes (two timesteps), whereas {\pFGW} has a higher interquartile range and mean. It shows that {\pFGW} performs the best at preserving the trajectory duration among the three approaches. 

We compute the standard deviation of {\CODm} and the linearity loss for trajectories that last for at least $45$ minutes (three timesteps); see the 2nd and 3rd columns of~\cref{fig:D1-statistics}. 
Three methods have similar performance in preserving the {\CODm} of cloud entities along the trajectory. 
On the other hand, {\pFGW} has a higher linearity loss compared to the other two methods. 
This is anticipated as cloud merging and splitting events have been observed to introduce undesirable shifts in the centroid position of the cloud system along the main trajectory. In particular, such position shifts can be drastic for large cloud systems, which are often long-lived for stratocumulus clouds.
In contrast, {\tobac} is less effective at identifying cloud merging and splitting events for large cloud entities (see~\cref{fig:D1-case-study} 4th column); {\PyFT} performs worse in maintaining the trajectory timespan. As a result, both tools generate fewer trajectories with high linearity loss.

\begin{figure*}[!ht]
\centering
\includegraphics[width=1.6\columnwidth]{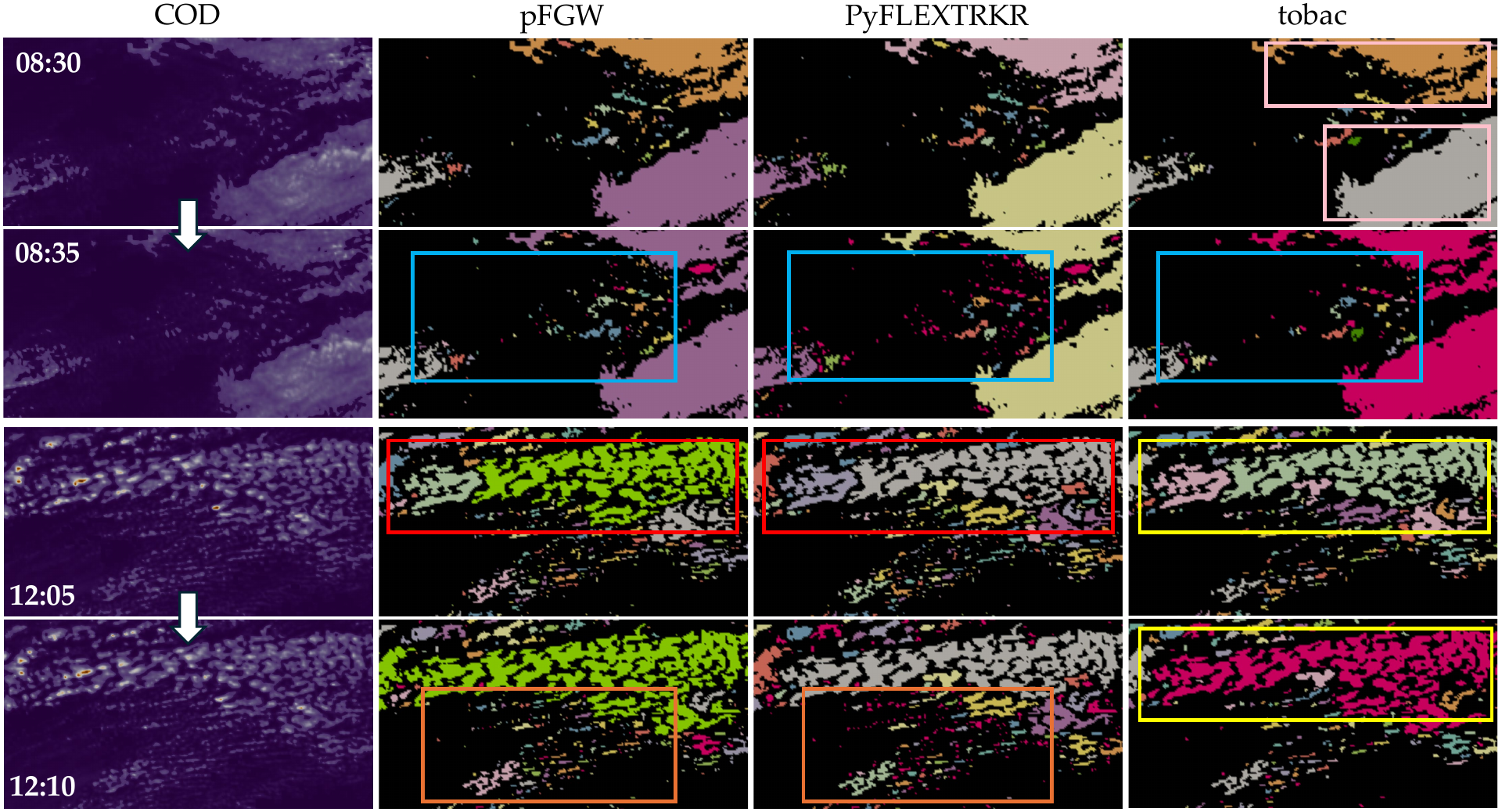}
\vspace{-2mm}
\caption{Tracking results of a region in the {\DL} dataset on May 1, 2018. From left to right: visualizations of COD fields, tracking results for {\pFGW}, {\PyFT}, and {\tobac}, respectively. 
The top two rows (resp. bottom two rows) are for the transition during the morning (resp. midday) period. 
All new cloud entities in the 2nd and 4th rows are colored magenta; others are colored by correspondences. 
Cyan boxes highlight the areas where {\PyFT} fails to track many small shallow cumulus clouds (shown as new entities in magenta). 
Yellow and pink boxes emphasize the suboptimal tracking results from {\tobac} when two large cloud objects merge.
}
\label{fig:D2-case-study}
\vspace{-6mm}
\end{figure*}

\subsection{Case Study: {\DL} Dataset}
\label{sec:result-d2}

The {\DL} dataset is collected above central Europe, where the complex land-driven convection strongly affects the low-level cloud systems. As the land warms up during the day, shallow cumulus often initiates in the morning, grows mature over time, and reaches its peak in the late afternoon. Hence, during its life cycle, the optical depth of shallow cumulus changes over time. 
Therefore, we divide the data into three periods for each day: 06:00 to 09:00 UTC for the \emph{morning}, 09:05 to 15:00 for the \emph{midday} period, and 15:05 to 17:55 for the \emph{late afternoon}. 
In particular, we are interested in the morning and midday periods, as these are typically when the initiation and maturation of shallow cumulus occur, respectively.

For the morning period, we set the superlevel set threshold to $9.0$, and for the midday period, we set it to $10.0$. This decision is based on the observed quality of cloud segmentation using superlevel set components, as outlined in~\cite{Fiolleau2013}, and the parameter sensitivity analysis described in~\cref{sec:method-detect-cloud-objects}, with further details in the supplement.
We do not simplify cloud entities by area because shallow cumulus clouds (particularly during the initiation stage) are smaller and have more gaps between individual cloud objects. 

\para{Cloud detection and tracking.}
We use the data from May 1, 2018, for our case study.
We check the transition from 08:30 to 08:35 UTC for data in the morning. For data in the midday, we check the transition from 12:05 to 12:10 UTC. 
We color all the new cloud entities in magenta at  8:35 UTC and 12:10 UTC, respectively. 
These new entities may arise from the formation of shallow cumulus, the splitting of a cloud entity, or the loss of cloud tracking.

The morning period reveals a cluster of shallow cumulus clouds developing near the center of the COD field, as illustrated in the first two rows of~\cref{fig:D2-case-study}. These shallow cumulus clouds are identified as a set of small cloud entities in the tracking results.
When comparing the performance of {\pFGW} and {\PyFT}, it becomes evident that {\PyFT} loses a significant number of trajectories for these tiny cloud entities; see the cyan box in the 3rd column. This limitation stems from the region-overlap-based approach used by {\PyFT} to track clouds. In small clouds, even slight positional shifts can lead to insufficient overlaps, causing the tracker to lose these clouds. 
In comparison, {\pFGW} is based on the clouds' geometric location and topological information, making it more robust when tracking small shallow cumulus clouds.

As the day progresses towards midday, the shallow cumulus system and many small cumulus cells evolve, growing thicker and merging into large cloud entities. In the bottom two rows, we observe large cloud entities in the top half of the image (red box), and clusters of small cloud entities in the bottom half (orange box). 
Both {\pFGW} and {\PyFT} exhibit similar performance in tracking these large cloud entities. 
However, for the small shallow cumulus, {\PyFT} generates numerous new cloud entities (see magenta cloud entities in the orange box), showing its limitations in tracking these smaller clouds consistently.
In comparison, {\pFGW} exhibits a better capability in consistently tracking small shallow cumulus clouds.

{\tobac} shows better performance in preserving the trajectories for small cloud objects than {\PyFT} (c.f. 3rd and 4th column).
However, we observe that the large cloud objects at 8:30 UTC  (1st row, pink boxes) are not tracked by {\tobac} during the morning period. These two objects merge into one at 8:35 UTC, which is treated as a new object by {\tobac}, similar to what we have observed in~\cref{sec:result-d1}.
Furthermore, {\tobac} also fails to track the large objects during the midday transition (3rd and 4th row, yellow box). 

\begin{figure}[!ht]
\centering
\vspace{-4mm}
\includegraphics[width=0.7\columnwidth]{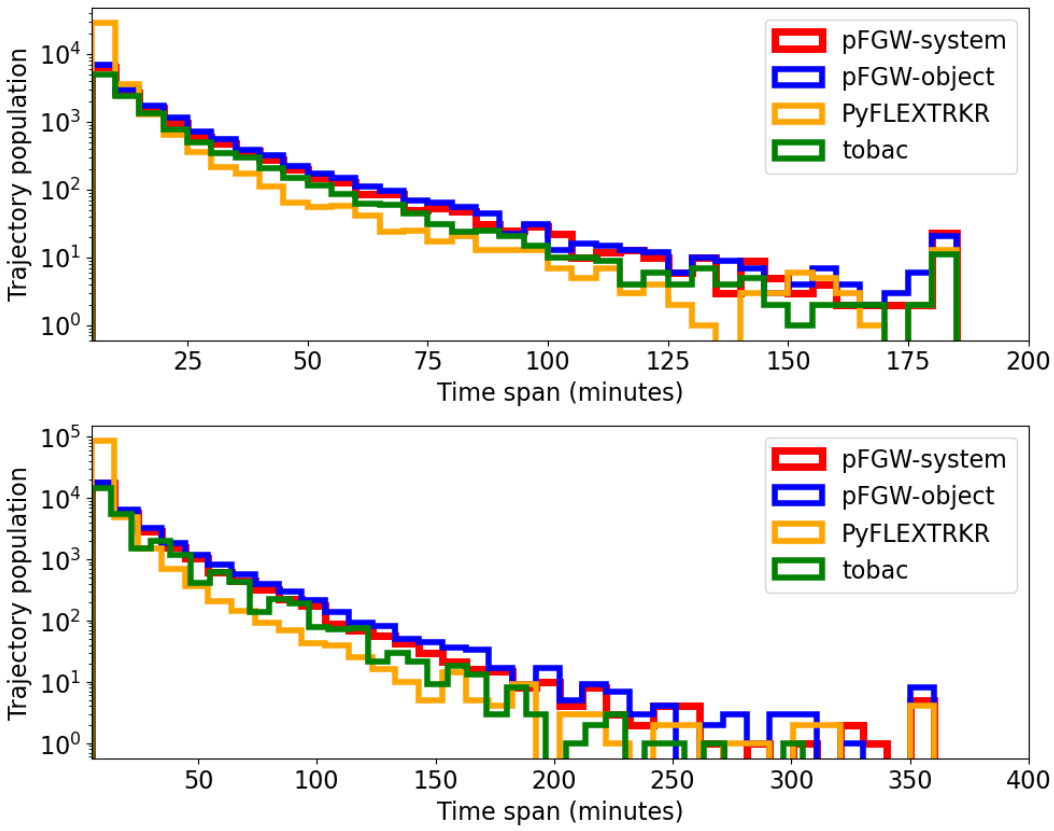}
\vspace{-3mm}
\caption{{\DL} dataset: distributions of trajectory timespans in log-scale for {\pFGW} tracking cloud systems (red) and objects (blue), {\PyFT} (orange), and {\tobac} (green); data aggregated over three days (May 1 and Jun 23 in 2018, and May 12 in 2019). The top row is for the morning period, and the bottom is for the midday.
}
\label{fig:D2-timespan}
\vspace{-2mm}
\end{figure}

\begin{figure}[!ht]
\centering
\vspace{-2mm}
\includegraphics[width=0.85\columnwidth]{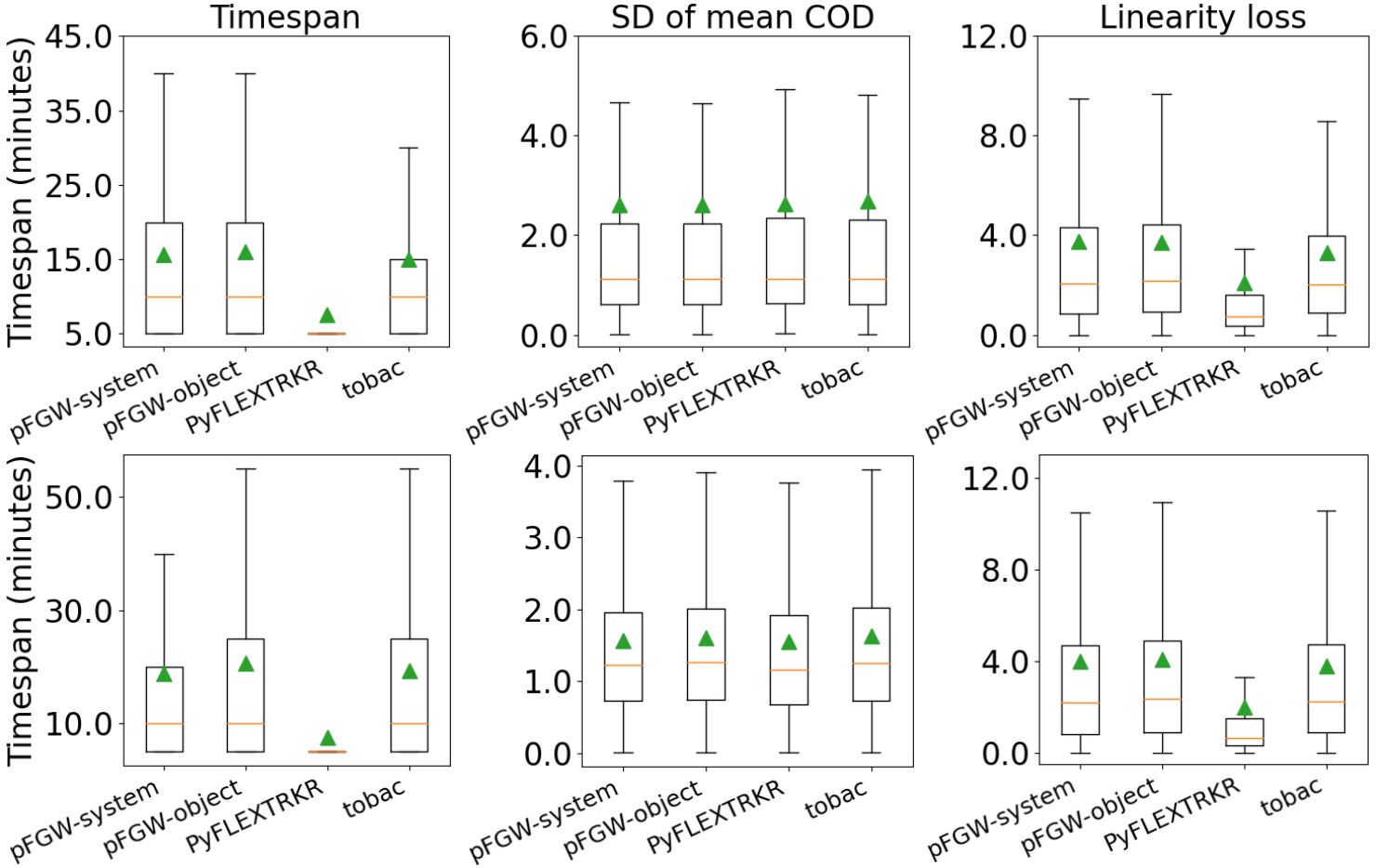}
\vspace{-3mm}
\caption{{\DL} dataset: statistical evaluation for the morning data (1st row, 06:00–09:00 UTC) and midday data (2nd row, 09:05–15:00 UTC). Box plots show the median (orange line), mean (green triangle), and interquartile range (box boundary) for three evaluation metrics.}
\label{fig:D2-statistics}
\vspace{-4mm}
\end{figure}

\para{Statistical evaluation.}
We perform the statistical evaluation similar to~\cref{sec:result-d1}.
We collect statistics for the datasets for three days on May 1, Jun 23, 2018, and May 12, 2019, respectively in~\cref{fig:D2-statistics}. The 1st row shows the distribution of evaluation metrics for the morning period, and the 2nd row shows the distribution for the midday period. 

We start with the trajectory timespan distribution. 
Among the three methods, {\PyFT} performs the worst on tracking small shallow cumulus, which constitutes the majority of the cloud entity population.  
Therefore, for both morning and midday periods, most trajectories from the {\PyFT} last for less than five minutes (one time step); see~\cref{fig:D2-timespan} and~\cref {fig:D2-statistics} 1st column.
Meanwhile, {\tobac} does not generate trajectories with a lifetime above $300$ minutes in the midday period as the other two methods do; see~\cref{fig:D2-timespan} 2nd row. 
This reflects our previous observation in~\cref{fig:D2-case-study} 4th column that {\tobac} performs worse than the other two methods in tracking large cloud entities, many of which are persistent in the {\DL} dataset.
In contrast, {\pFGW} performs the best on maintaining trajectories for small shallow cumulus in the morning and has a similar trajectory lifetime distribution to {\tobac} in tracking cloud objects during the midday period; see~\cref{fig:D2-statistics} 1st column. 
By merging nearby cloud objects into cloud systems, {\pFGW} may get fewer trajectories with a long lifetime. However, {\pFGW} still performs better than {\tobac} in maintaining long-term trajectories in the midday period; see~\cref{fig:D2-timespan} 2nd row. 

We compute the standard deviation of {\CODm} and the linearity loss for trajectories that last for at least 15 minutes (three timesteps). 
When evaluating the ability to preserve {\CODm} along the same trajectory, all three methods have comparable performances during both morning and midday; see the 2nd column of~\cref{fig:D2-statistics}.

Lastly, because there are fewer large cloud entities in the {\DL} dataset, the linearity loss of {\pFGW} trajectories is similar to that of {\tobac}. On the other hand, it is anticipated that {\PyFT} generates trajectories with the least linearity loss because {\PyFT} often loses track of cloud entities on the {\DL} dataset. 

\subsection{Comparison with Topology-based Tracking Tools}
\label{sec:result-topology-tools}
For completeness, we further compare {\pFGW} against two topology-based general-purpose tracking tools: the Lifted Wasserstein Matcher ({\LWM})~\cite{SolerPlainchaultConche2018} and the Wasserstein distance between merge trees ({\MTW})~\cite{PontVidalDelon2021}. 
{\LWM} extends the Wasserstein distance between persistence diagrams by incorporating critical point locations. It aims to solve a minimum-cost matching problem between persistence diagrams, where the cost of matching is a linear combination of the Wasserstein distance and the Euclidean distance between the matched critical points. Additionally, the cost of adding or removing a persistence pair is a linear combination of its persistence and the Euclidean distance between its two critical points. In contrast, the {\MTW} framework extends the edit distance between merge trees by introducing constraints. It computes the edit distance between branch decomposition trees generated from merge trees, with the matching cost based on the persistence of branches. However, this framework does not consider geometric information when tracking features.

\para{Experimental settings.} 
For {\pFGW}, we use an Euclidean distance threshold at $28$km (see the supplement for a discussion); our parameter tuning for $m$ guarantees that all matchings between anchor points beyond this distance threshold are ignored. 
The parameter settings of {\LWM} focus on the weight balancing the Wasserstein distance between persistence pairs and the Euclidean distance between critical points. 
Building on previous works~\cite{SolerPlainchaultConche2018, LiYanYan2023}, we normalize the range of the two distances and set the weight of the Euclidean distance to \( 1.0 \). We then gradually increase the weight of the Wasserstein distance, denoted as \( \beta \), from \( 0 \) to assess the impact of persistence information. 
For {\MTW}, we follow the recommended parameter setting from its original work~\cite{PontVidalDelon2021}. 

We compute the cloud system trajectories for {\LWM} and {\MTW} using a postprocessing pipeline similar to that of {\pFGW}. We compare the cloud system tracking performance using the statistics described in~\cref{sec:evaluation-metric}.
Additional experimental details are in the supplement.

In the following, we compare the anchor point matching results from all three topology-based approaches using the {\DM} dataset. 
Specifically, for {\LWM}, we report the results using $\beta \in \{0.0, 0.1, 0.2\}$ to examine the impact of persistence diagram information on tracking.
$\beta=0$ means that the cost function of {\LWM} uses all geometric information. Increasing $\beta$ adds the weight of the Wasserstein distance in the cost function. When $\beta=1.0$, we balance the weight between the Wasserstein distance and the Euclidean distance in the cost function.
Furthermore, we perform the same statistical evaluation process in~\cref{sec:evaluation-metric} to compare the cloud system tracking results using the {\DM} dataset. 
Additional comparisons are in the supplement.

\begin{figure}[!ht]
\vspace{-4mm}
\centering
\includegraphics[width=0.9\columnwidth]{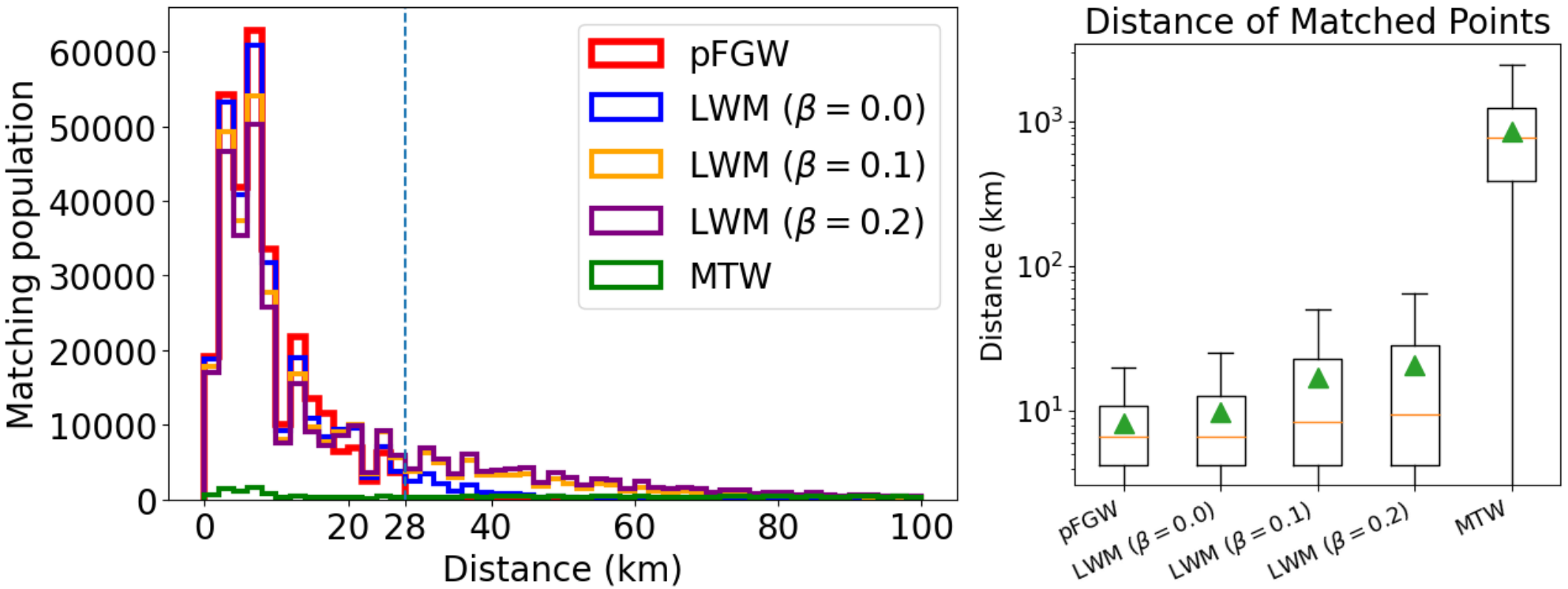}
\vspace{-2mm}
\caption{{\DM} dataset: histograms (left) and box plots in log-scale (right) for distributions of the Euclidean distances between matched anchor points. The Euclidean distance threshold for {\pFGW} is $28$km, as highlighted by the vertical dotted line in the histogram. }
\label{fig:topotools-distance}
\vspace{-3mm}
\end{figure}

\para{Anchor point matching.}
We examine the distribution of the Euclidean distance between matched anchor points in~\cref{fig:topotools-distance}. The left histograms show that the distributions of Euclidean distances are similar between {\pFGW} and {\LWM}, with some differences on the right tail. 
The Euclidean distances between all matched points for {\pFGW} are below $28$km.
In comparison, there are matched nodes in the results of {\LWM} with $\beta=0$ with the Euclidean distance beyond $28$km, which is less likely to happen due to physical constraints~\cite{heikenfeld_2019}.
This is because the cost of matching points to the diagonal is fixed by the saddle-maxima relation of the persistence pair, losing the flexibility to adapt to what the application needs.
As the $\beta$ value of {\LWM} increases to $0.1$ and then to $0.2$, the number of matched nodes with short Euclidean distances decreases, and the number of faraway matched nodes increases; see~\cref{fig:topotools-distance} left. 
In addition, in~\cref{fig:topotools-distance} right, the mean and median Euclidean distance between matched nodes for {\LWM} increases as $\beta$ increases from $0$ to $0.2$.
These observations indicate that adding the weight of the Wasserstein distance does not benefit the anchor point tracking.
In contrast to {\LWM}, {\pFGW} uses the tree distance between anchor points to encode the topological information, reflecting the locality of anchor points within cloud objects. Such information is more robust than the persistence diagram information when performing feature-tracking tasks in such complex datasets.
Among the three approaches, {\MTW} performs the worst. Without geometric location information, {\MTW} does not find many matchings between nearby anchor points. The mean and median Euclidean distance between matched anchor points is much higher than the other two methods; see~\cref{fig:topotools-distance} right.

\begin{figure}[!ht]
\centering
\vspace{-3mm}
\includegraphics[width=0.9\columnwidth]{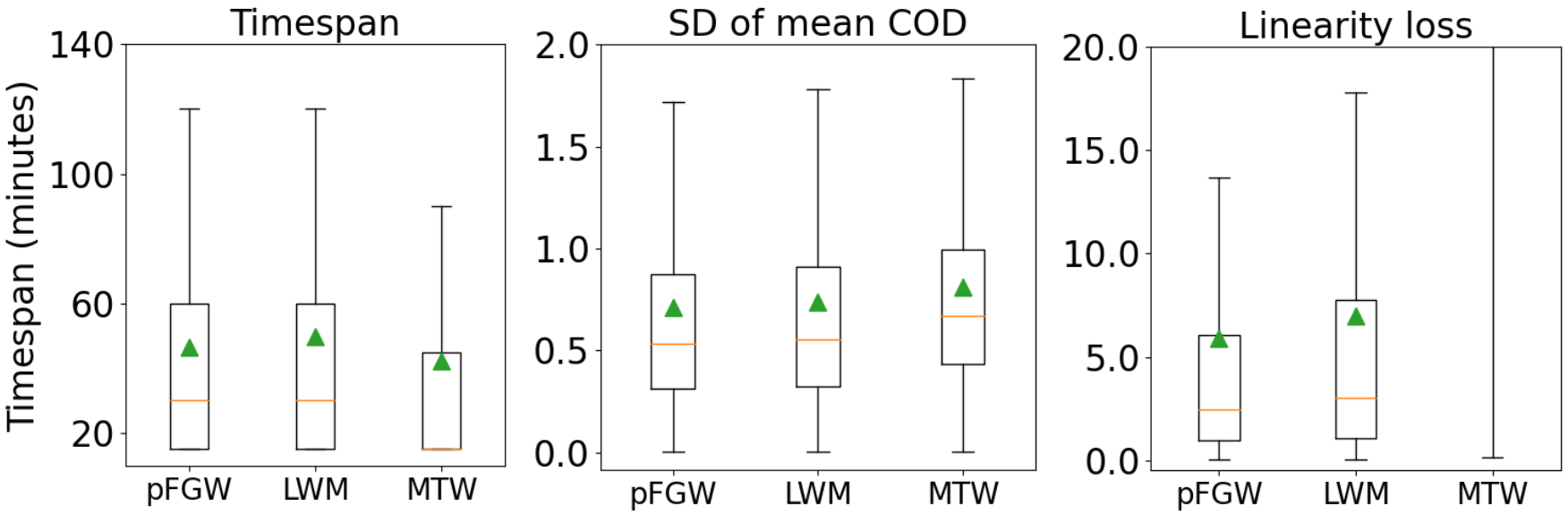}
\vspace{-3mm}
\caption{{\DM} dataset: box plots showing the median (orange line), mean (green triangle), and interquartile range (box boundary) of the distribution for three topology-based tracking methods. The box for the linearity loss for {\MTW} exceeds the plot's upper bound.}
\label{fig:topotools-statistics}
\vspace{-5mm}
\end{figure}

\para{Statistical evaulation.}
\cref{fig:topotools-statistics} shows the distribution of  statistics for three topology-based methods. Specifically, for {\LWM}, we fix $\beta=0$ because it  performs the best for anchor point matching among all choices of $\beta$. 
We start with the trajectory timespan in~\cref{fig:topotools-statistics} left. Among the three methods, {\MTW} performs the worst in maintaining trajectory continuity. The mean and median trajectory timespan for {\MTW} tracking results are the lowest. 
In comparison, {\pFGW} and {\LWM} have similar performance, while {\LWM} has a slightly higher mean timespan for trajectories. 
For the standard deviation of mean COD, all three approaches demonstrate similar distributions for their results. 
This indicates that all three methods perform similarly in matching cloud systems with similar COD distributions, which are partly reflected by the anchor point COD values.
Lastly, {\pFGW} performs the best in minimizing the linearity error, {\LWM} the second, and {\MTW} much worse than the other two; see~\cref{fig:topotools-statistics} right (where the boxplot of MTW goes beyond the boundary). This is expected because {\MTW} performs poorly in matching nearby anchor points. 
In addition,~\cref{fig:topotools-distance} shows that {\LWM} produces more matchings between distant anchor points compared to {\pFGW}, resulting in higher trajectory linearity loss for {\LWM}.

%% file: sec-evaluation.tex

%% file: sec-discussion.tex
\section{Conclusion and Discussion}
\label{sec:discussion}

The case studies and statistical evaluations provide several important takeaways.
First, our framework operates with cloud systems instead of cloud objects, reducing sensitivity to threshold selection and producing fewer short-lived cloud trajectories. Tracking low-level clouds as systems offers deeper insights into their proximity and evolution.
Second, our framework demonstrates strong performance in tracking clouds compared to two state-of-the-art cloud tracking methods. Notably, it is the most consistent in tracking small shallow cumulus clouds over land as well as large stratocumulus over the ocean.
In future work, we aim to enhance tracking quality by integrating additional cloud variables, such as cloud fraction, cloud liquid water path, and cloud top height~\cite{freeman_2024}.

%% file: sec-cloud-data.tex
\section{Cloud Data}
\label{sec:cloud-data}

Geostationary satellites offer continuous measurements of cloud systems as they evolve over time, a capability utilized in cloud remote sensing since the launch of the first Applications Technology Satellite (ATS-1) in 1966~\cite{menzel_2001}. 
This study employs two distinct resolutions of cloud optical depth (COD) retrievals~\cite{Roebeling2006, deneke_2021, werner_increasing_2020} derived from the visible and near-infrared channels of the SEVIRI instrument onboard Meteosat's second-generation satellites.

\begin{figure}[!ht]
\centering
\vspace{-2mm}
\includegraphics[width=0.8\columnwidth]{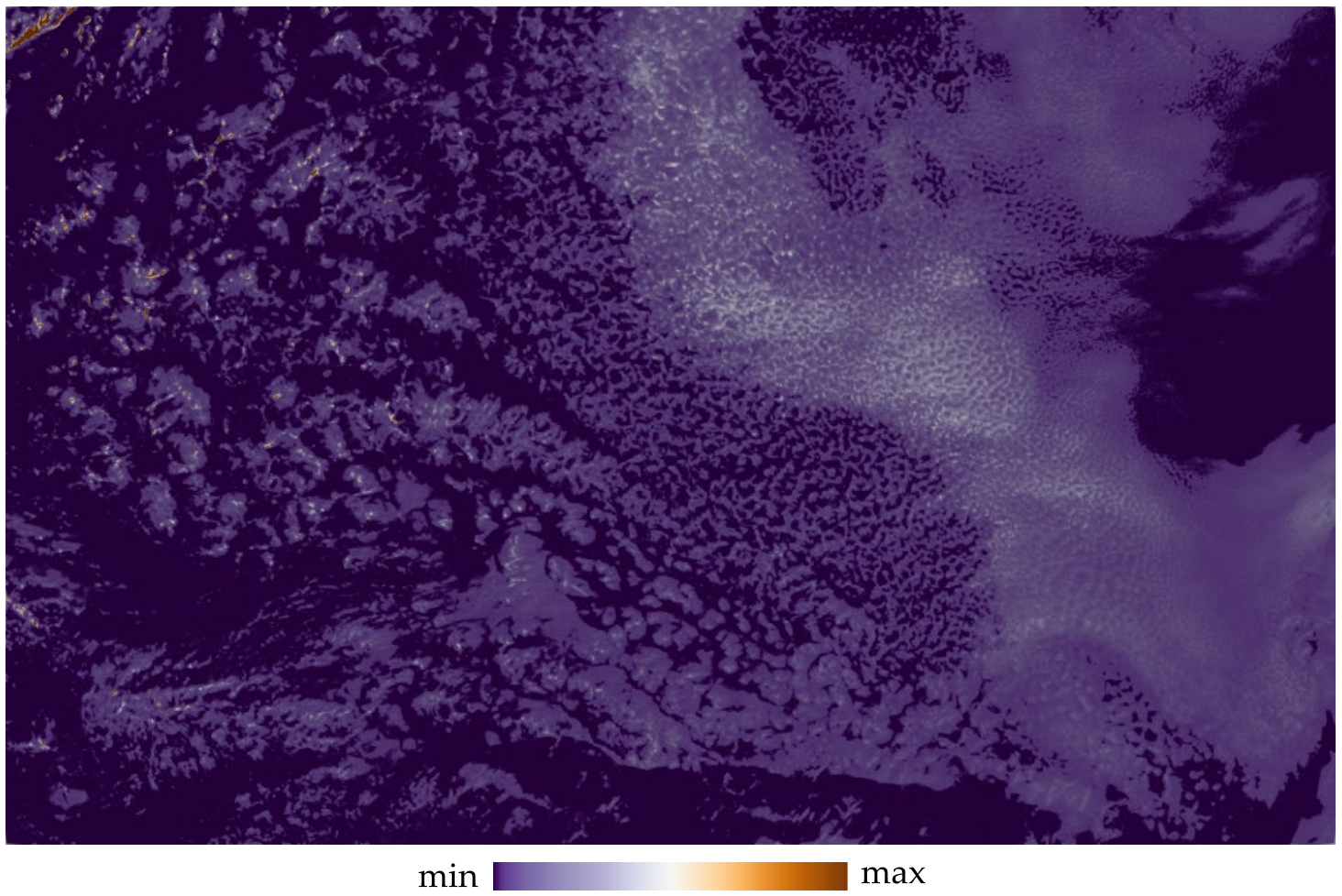}
\vspace{-2mm}
\caption{A snapshot of the COD field at 09:00 UTC on Aug 1, 2023, from a marine stratocumulus cloud dataset over the ocean west of Africa.}
\label{fig:cod-field}
\vspace{-2mm}
\end{figure}

The first dataset, referred to as {\DM}, utilizes the CLAAS-3.0 product \cite{Meirink_2022}, focusing on marine stratocumulus clouds over the ocean west of Africa (26.74°S to 4.52°S, 10.52°E to 27.99°W) during August 2023. This dataset retains SEVIRI's native resolution, with a 15-minute temporal repeat cycle and a spatial resolution of 3 km at nadir. Stratocumulus cloud systems are prevalent in this region during the austral winter months (July–September)~\cite{klein_1994, schween_2022}.
~\cref{fig:cod-field} shows an example of the COD field from the {\DM} dataset.

The second dataset, labeled as {\DL}, covers shallow cumulus cloud systems over central Europe (47.3°N to 55.4°N, 1.9°E to 9.3°E) across 13 selected days of low-level cloud occurrences from April to September between 2018 and 2019. This dataset simulates the capabilities of the Meteosat Third Generation (MTG) mission, offering an enhanced spatial resolution of 2 km $\times$ 1 km and a 5-minute temporal repeat cycle. {\DL} significantly improves the standard MSG cloud dataset, enabling more detailed tracking of convective systems in central Europe~\cite{deneke_2021, werner_increasing_2020}.    

In the visible range, satellites receive part of the solar radiation reflected by the clouds or the earth’s surface. The near-infrared channels going beyond the visible range help examine how objects reflect, transmit, and absorb the sun's infrared emission. The retrieval algorithm operates on the principle that cloud reflectance is predominantly governed by COD with minimal sensitivity to particle size at visible wavelengths. In contrast, at near-infrared wavelengths, cloud reflectance is primarily influenced by particle size~\cite{Nakajima1990, han_1994, King_2004}.

Modern retrieval methods, such as the Cloud Physical Properties (CPP) algorithm used with SEVIRI by the Royal Meteorological Institute of the Netherlands \cite{Roebeling2006}, rely on a combination of non-absorbing visible wavelengths (0.6 or 0.8 µm) and near-infrared wavelengths (1.6 or 3.8 µm). While the 1.6 µm wavelength is well-suited for thicker clouds, the 3.8 µm wavelength is better for thinner clouds. However, retrievals at 3.8 µm involve greater uncertainty due to their proximity to thermally emitted radiance and the lower solar irradiance at this wavelength compared to 1.6 µm. Consequently, the CPP algorithm primarily utilizes 0.6 and 1.6 µm reflectances for retrieving COD, particle size, and cloud liquid water path (CLWP).

We chose COD as the input for our cloud system tracking framework due to the following reasons: (i) COD eliminates the need to account for solar zenith angle and surface characteristics, both of which significantly impact reflectance values, and (ii) COD is the most accurate retrieval product available at the enhanced resolution~\cite{deneke_2021}. However, one must note that the COD uncertainties become large outside a value range of 5 - 50 but remain below 8-10\% within this range~\cite{Platnick2017}.

%% file: sec-implementation.tex
\section{Implementation}
\label{sec:implementation}

All experiments are done on a laptop with a 12th Gen Intel(R) Core(TM) i9-12900H 2.50 GHz CPU with 32 GB memory. 
We use the python library \textit{scipy}~\cite{Virtanen2020SciPy-NMeth} to identify superlevel set components for cloud object detection.
We use the \textit{ParaView 5.11.1}~\cite{AhrensGeveciLaw2005} and \textit{TTK 1.1.0}~\cite{TiernyFavelierLevine2018} to compute merge trees and topological zones.
We follow the work of Li et al.~\cite{LiYanYan2023} to compute the pFGW distance for merge tree matching, which has open-source code on GitHub~\cite{LiYanYan2023gwmt}.
We will provide our implementation on GitHub upon publication.

%% file: sec-parameters.tex
\section{Experimental Parameters}
\label{sec:parameters}

We discuss experimental parameters in addition to those mentioned in Sec.~5 and Sec.~6.

\subsection{Cloud Object Detection and Simplification}

\begin{figure}[!ht]
\centering
\includegraphics[width=1.0\columnwidth]{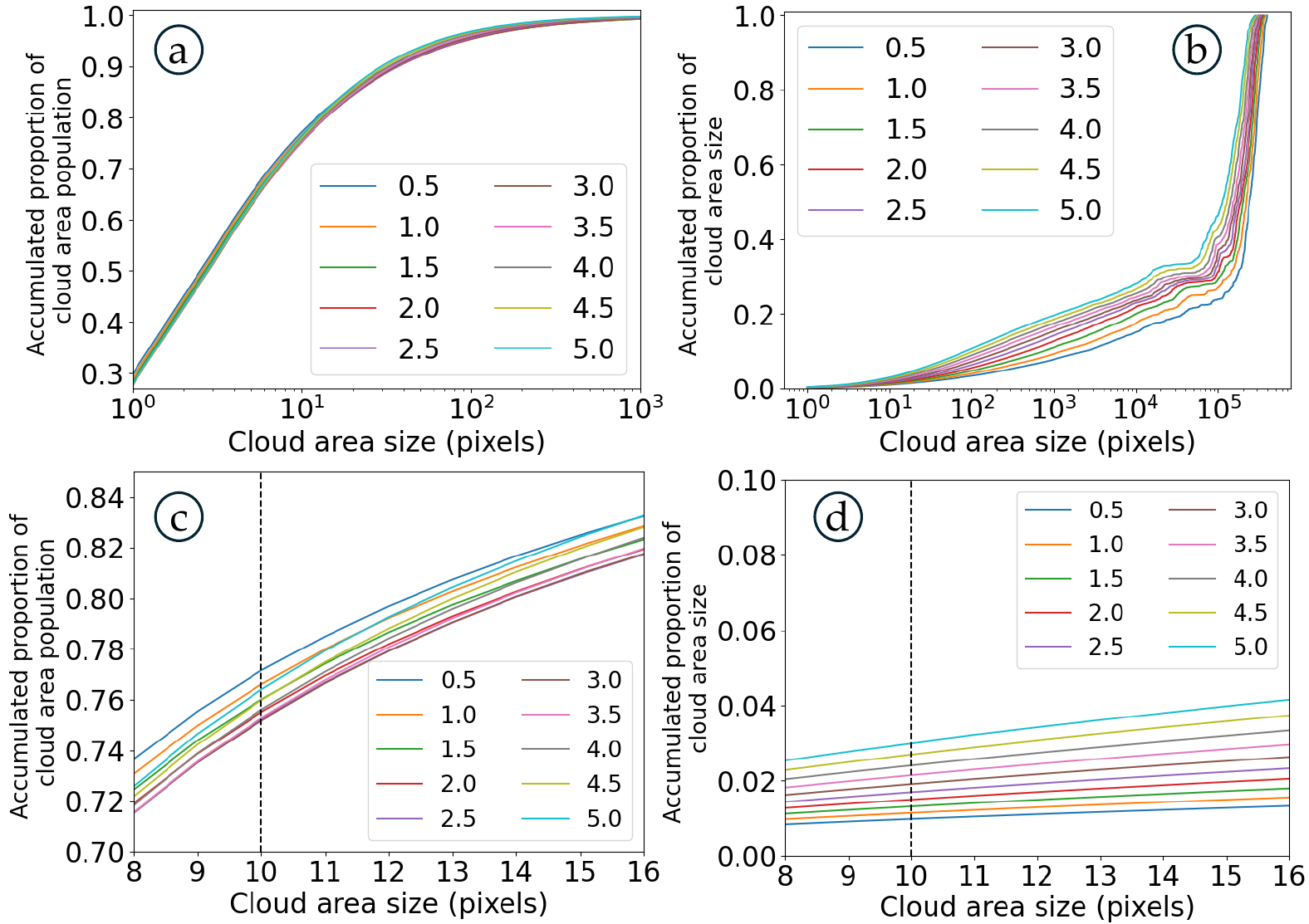}
\vspace{-6mm}
\caption{(a) the accumulative proportion of cloud object number density. (b) the accumulative proportion of cloud area size. Each curve represents a superlevel set threshold value between $0.5$ and $5.0$. (c) and (d) are the zoom-in views for (a) and (b), respectively. Notice that the x-axes in (a) and (b) are based on log-scale, whereas those in (c) and (d) are not. Statistics are computed on the {\DM} dataset from Aug 1 to Aug 14, 2023.}
\label{fig:cloud-area-statistics}
\vspace{-4mm}
\end{figure}

\para{Object detection parameter sensitivity analysis.} Currently, there is no consensus on the threshold value $a$ to detect low-level clouds. Following established practices~\cite{heikenfeld_2019}, we test a range of thresholds from $0.5$ to $5.0$ at a gap of $0.5$ to analyze the impact of the threshold on the cloud area size and numbers. 
This approach is similar to using brightness temperature thresholds for tracking deep convective cloud systems~\cite{heikenfeld_2019} and references therein. 
We use the statistics computed on the {\DM} dataset as an example.
\cref{fig:cloud-area-statistics}(a) shows the cumulative distribution of cloud object number density as cloud size increases. 
The distribution curves appear consistent across different superlevel set thresholds, indicating that the cloud area size distribution is not highly dependent on the superlevel set threshold within the range of $[0.5, 5.0]$.
In particular, more than $70\%$ of cloud objects are below $10$ pixels regardless of the threshold; see the zoom-in view at~\cref{fig:cloud-area-statistics}(c).
In~\cref{fig:cloud-area-statistics}(b), the differences among curves are more noticeable. 
For example, when the threshold is $0.5$, cloud objects with more than $1000$ pixels contribute to $92.15\%$ of the total cloud coverage. 
In comparison, $80.36\%$ of the cloud area coverage comes from clouds with more than $1000$ pixels when the threshold is $5.0$. 
This result indicates that when choosing a threshold that is too small, large clouds will dominate the cloud area coverage, and we may mistakenly interpret multiple cloud systems as a single one altogether.
On the other hand, if we use a very high threshold, we may obtain too many small cloud objects over-segmented from a large one. 
Therefore, we choose a threshold of $2.0$ to avoid potential issues with extreme values. 

The {\DL} dataset exhibits more complex characteristics and patterns of low-level clouds. Specifically, we can observe the initiation and maturation process for shallow cumulus clouds at different times of the day due to the overland convection. The variation of COD for low-level clouds has to be considered when choosing the superlevel set thresholds. 
We perform parameter sensitivity analysis for two time periods: 06:00 to 09:00 UTC for the morning and 09:05 to 15:00 UTC for the midday. 

\begin{figure}[!ht]
\centering
\includegraphics[width=1.0\columnwidth]{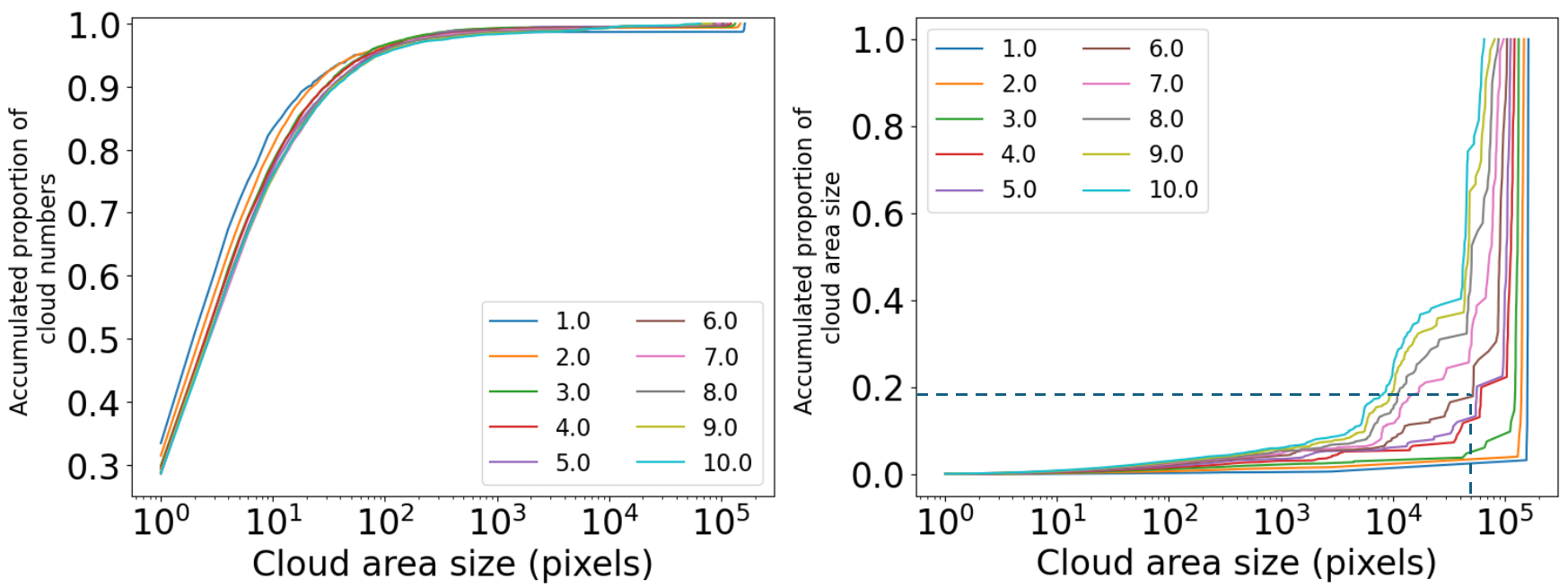}
\vspace{-6mm}
\caption{(a) the accumulative proportion of cloud object number density. (b) the accumulative proportion of cloud area size. Each curve represents a superlevel set threshold value between $1.0$ and $10.0$. Statistics are computed on the {\DL} dataset, including the snapshots collected between 06:00 and 09:00 UTC for all $13$ selected days between $2018$ and $2019$. Notice that the x-axes are based on log-scale.}
\label{fig:D2-parameter-sensitivity-morning}
\vspace{-2mm}
\end{figure}

\begin{figure}[!ht]
\centering
\includegraphics[width=1.0\columnwidth]{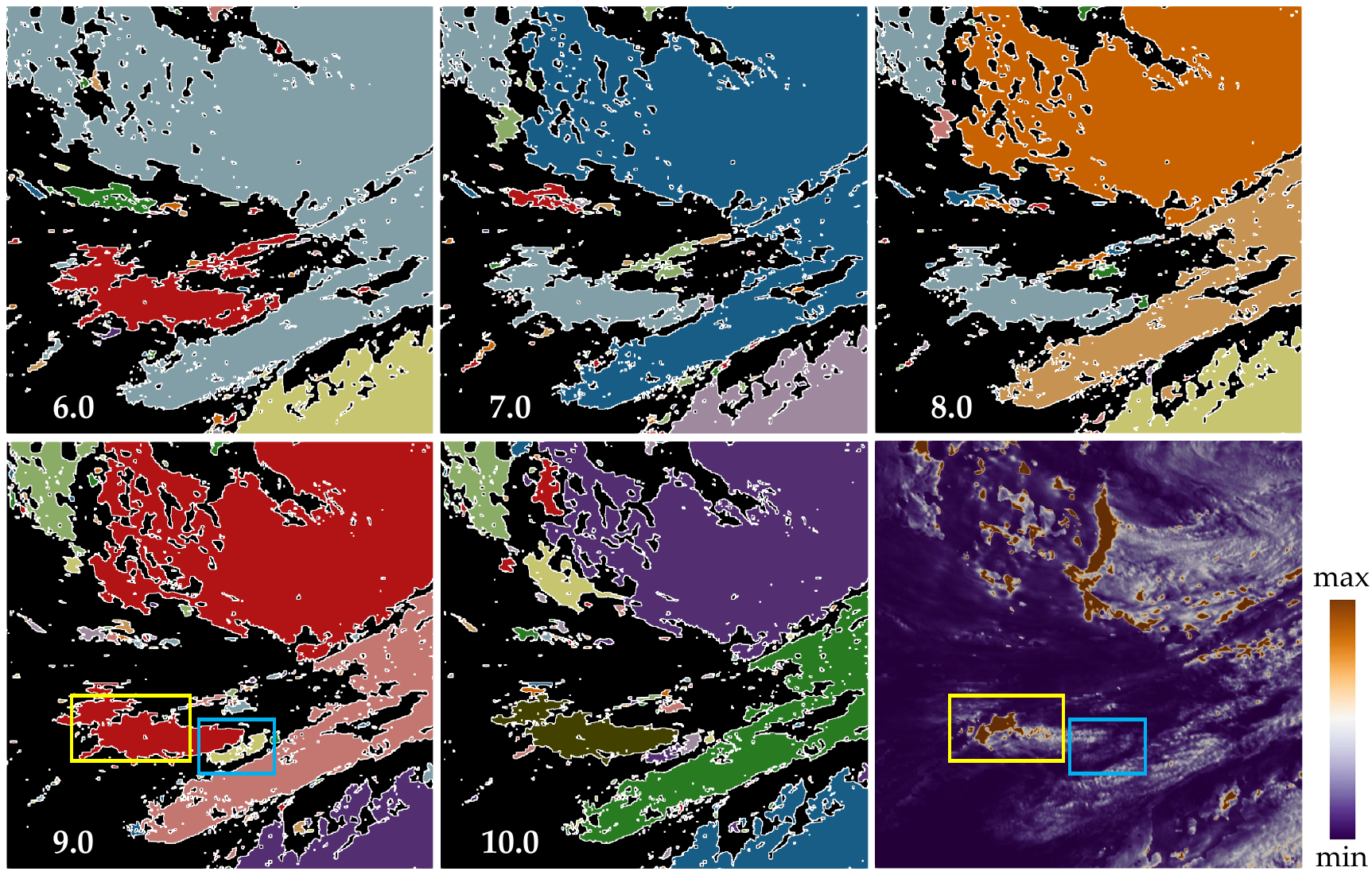}
\vspace{-6mm}
\caption{Individual cloud objects detected using superlevel set thresholds between $6.0$ and $10.0$ for the COD field data (2nd row, 3rd column) collected on May 1, 2018, at 07:00 UTC. The white numbers represent the superlevel set thresholds. Regions within the blue boxes and yellow boxes are identified as a single cloud object until the superlevel set threshold reaches $9.0$.}
\label{fig:D2-threshold-justification-example-morning}
\vspace{-2mm}
\end{figure}

We test the threshold between $1.0$ and $10.0$ to perform a parameter sensitivity analysis similar to the {\DM} dataset for the morning data.~\cref{fig:D2-parameter-sensitivity-morning} shows the accumulative proportion of cloud object number density and cloud area coverage. The dotted lines in~\cref{fig:D2-parameter-sensitivity-morning}(b) highlight a point on the curve for the threshold at $6.0$, indicating that less than $80\%$ of the low cloudiness is due to cloud clusters smaller than $50,000$ pixels. 
When the threshold is below $6.0$, this percentage becomes lower. This observation indicates that it may be difficult to identify small shallow cumulus clouds from the large cloud cluster when the threshold is low. 
We further examine the cloud detection results using different superlevel set thresholds. 
In these results, we select the lowest threshold that reasonably separates cloud objects.~\cref{fig:D2-threshold-justification-example-morning} shows an example of cloud detection results with the threshold between $6.0$ and $10.0$. The COD field in the blue box has lower COD values compared to the high-value area in the yellow box. The area in the blue and yellow boxes is identified as a single cloud object until the threshold reaches $9.0$. Therefore, we choose $9.0$ as the threshold for this snapshot.

\begin{figure}[!ht]
\centering
\includegraphics[width=1.0\columnwidth]{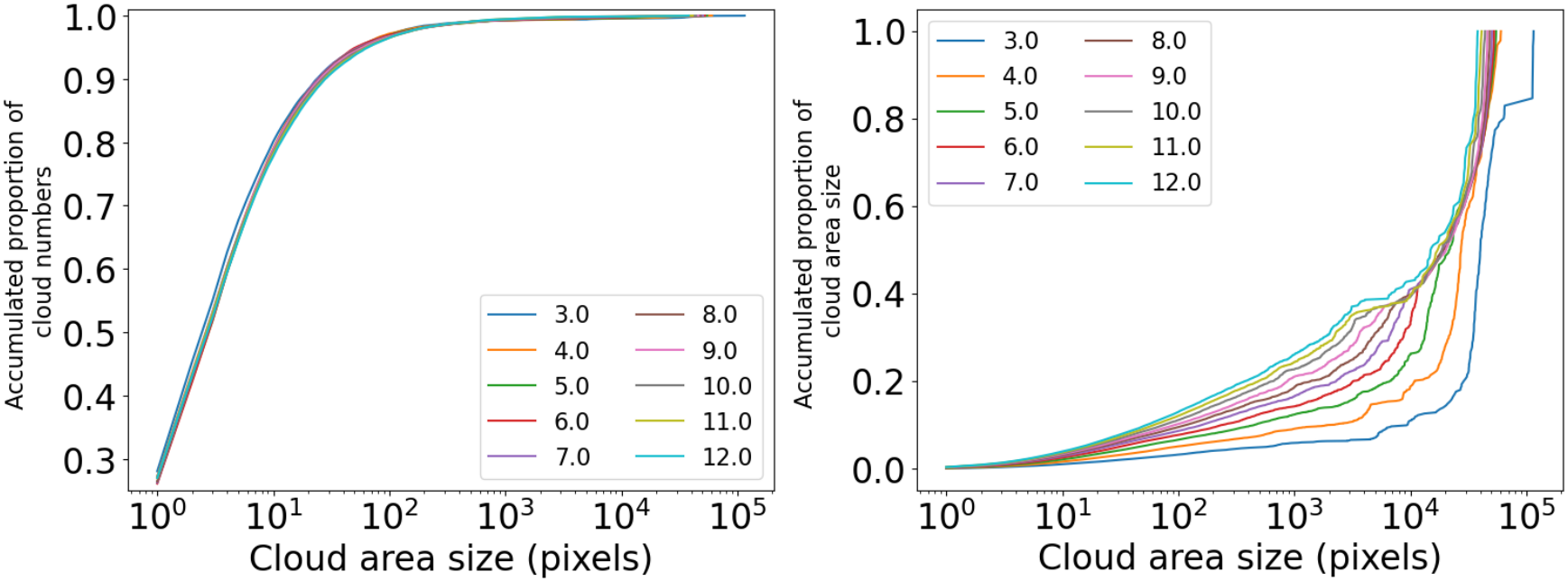}
\vspace{-6mm}
\caption{(a) the accumulative proportion of cloud object number density. (b) the accumulative proportion of cloud area size. Each curve represents a superlevel set threshold value between $1.0$ and $10.0$. Statistics are computed on the {\DL} dataset, including the snapshots collected between 09:05 and 15:00 for all $13$ selected days between $2018$ and $2019$.}
\label{fig:D2-parameter-sensitivity-midday}
\vspace{-4mm}
\end{figure}

\begin{figure}[!ht]
\centering
\includegraphics[width=1.0\columnwidth]{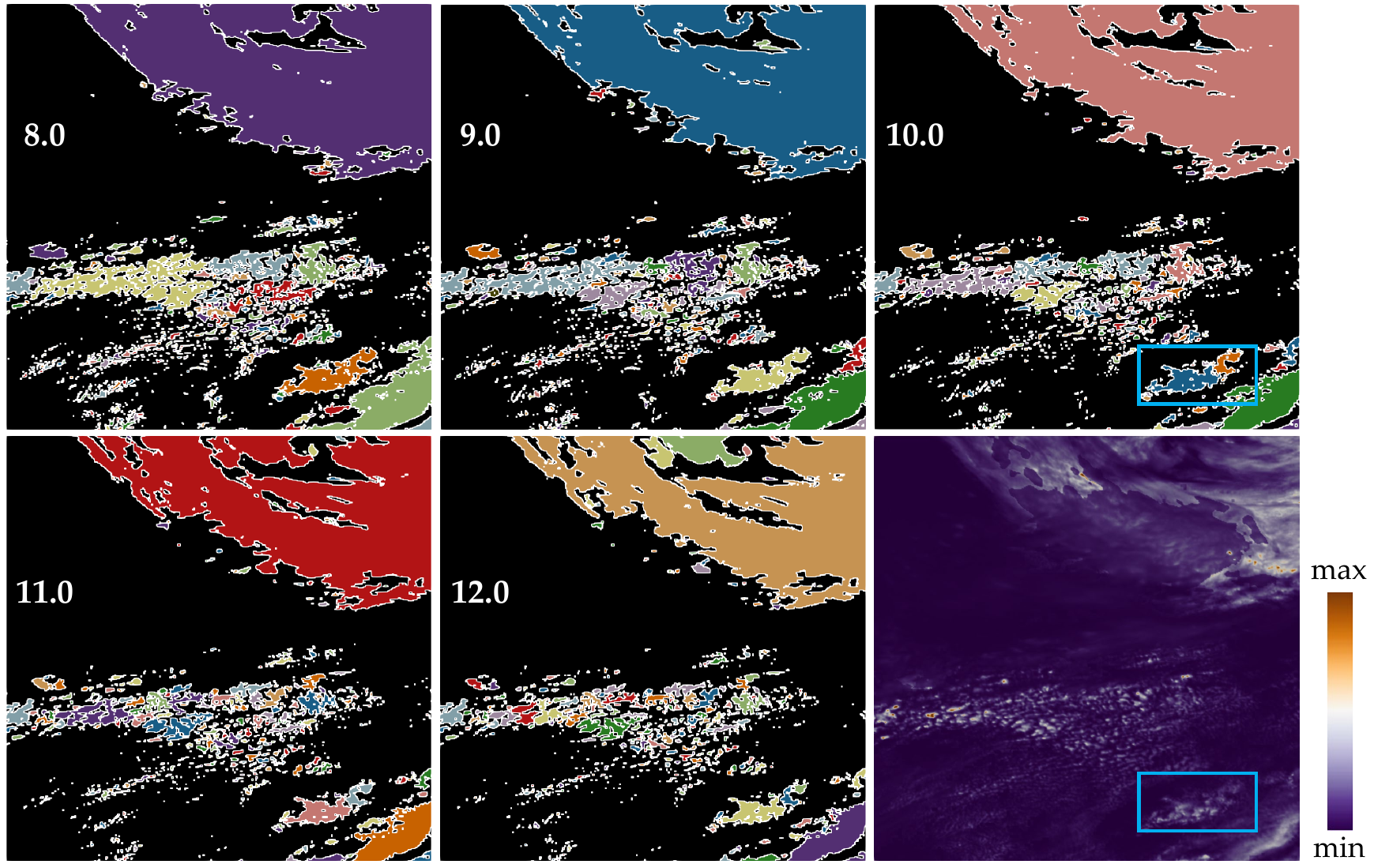}
\vspace{-6mm}
\caption{Individual cloud objects using superlevel set thresholds between $8.0$ and $12.0$ for the bottom right COD field data (2nd row, 3rd column) collected on May 1, 2018, at 11:00 UTC. The white numbers represent the superlevel set thresholds.}
\label{fig:D2-threshold-justification-example-midday}
\vspace{-2mm}
\end{figure}

\cref{fig:D2-parameter-sensitivity-midday} shows the cumulative distribution of cloud number density and cloud area coverage for the midday data in {\DL}. 
We change the range of thresholds for testing to $[3.0, 12.0]$ because we expect to see shallow cumulus clouds in higher COD values during midday.
Specifically, the cloud area size distribution (see~\cref{fig:D2-parameter-sensitivity-midday} right) is sensitive to the superlevel set threshold below $6.0$. We examine the snapshots using the threshold from $6.0$ to $12.0$. 
~\cref{fig:D2-threshold-justification-example-midday} shows an example COD field during the midday period, with a subset of cloud detection results using the threshold between $8.0$ and $12.0$. The COD field within the blue box in~\cref{fig:D2-threshold-justification-example-midday} has higher values on the bottom left and lower values on the top right. $10.0$ is the lowest threshold value that separates these two regions as individual objects. Therefore, we use $10.0$ as the threshold for this snapshot.

We examine snapshots at 07:00, 08:50, 11:00, and 13:00 UTC for each day in the {\DL} dataset and determine the best threshold for each snapshot. We summarize the decisions for snapshots at 07:00 and 08:50 UTC for the morning data and those for snapshots at 11:00 and 13:00 UTC for the midday data. 
The final threshold decision is made by choosing the most popular threshold for the time period.

\para{Object simplification parameter tuning.}
We refer to the statistics of cloud area size in~\cref{fig:cloud-area-statistics} to determine the area-based simplification level for the dataset of stratocumulus: by ignoring cloud objects smaller than $10$ pixels, we can get rid of more than $70\%$ of cloud objects from the field (see~\cref{fig:cloud-area-statistics}(c)), and still cover more than $97\%$ of the total cloud area (see~\cref{fig:cloud-area-statistics}(d)), regardless of the superlevel set threshold. Therefore, we choose the area-based simplification level to be 10 pixels.

\subsection{Cloud Tracking}

\label{sec:parameters-tracking}
\para{Parameter tuning for pFGW distance.} 
Following~\cite{LiYanYan2023}, to compute the pFGW distance, we need to tune the two parameters: $\alpha$ and $m$.
Recall that $\alpha$ is the parameter balancing the weight between the intrinsic information of merge trees and the extrinsic information of nodes.
$\alpha \rightarrow 0$ leads to a high weight on keeping the geometric locations of nodes in tracking, whereas $\alpha \rightarrow 1$ leads to a high weight on keeping the merge tree structure. 
$\alpha=0.5$ puts the same weight on both parts.
We attempt $\alpha \in \{0.2, 0.4\}$, putting a higher weight on preserving the geometric location of nodes while still considering the intrinsic merge tree structure. We choose the better results between the two choices of $\alpha$ for demonstration.

The parameter $m$ determines the amount of probability mass to be preserved in the partial optimal transport, allowing feature appearance and disappearance.
We follow a similar strategy to~\cite{LiYanYan2023} to adjust the parameter $m$. 
In particular, we determine the threshold for the maximum possible distance between matched nodes from adjacent time steps (we discuss this option in the next paragraph).
Then, we search for the highest $m$ value such that no matching between nodes farther than the above threshold exists. 
We want to keep the highest probability in the coupling without introducing obviously incorrect matching.

We noticed that the better choice of $\alpha$ is $0.4$ for {\DM}. This is because for stratocumulus, there are often a few anchor points inside the cloud objects due to their relatively large area. Using a higher $\alpha$ value to emphasize the locality of anchor points within each cloud object is beneficial. 
On the other hand, the better $\alpha$ for {\DL} is $0.2$. This is because many tiny cloud objects in {\DL} only have one anchor point, which means that there is less anchor point locality information than {\DM}. It is necessary to add the weight of geometric location information in the loss function to obtain accurate tracking.

\para{Maximum matched distance.}
Both {\tobac} and {\pFGW} have a parameter of the maximum distance between two matched clouds at adjacent time steps. 
For {\tobac}, the parameter helps restrain the size of the neighborhood from searching within at the next time step. For {\pFGW}, this parameter helps determine the parameter $m$, the probability mass (i.e., the amount of ``goods'') to be preserved in the partial optimal transport (see Sec. 5.3).
Because the time interval between snapshots and the domain area is consistent for a dataset, we transform this threshold into a limit of the average speed of clouds.
For the {\DM} dataset, we use the distance threshold corresponding to an average moving speed at $20m/s$ as the highest translation speed of feature points from {\tobac}, following the analysis in~\cite{heikenfeld_2019}.
For anchor points from {\pFGW}, this average speed threshold is $30m/s$ because the anchor point position is more sensitive to COD values.  
For the {\DL} dataset, we increase the average speed threshold $40m/s$ to adapt to more complex low-level cloud changes for both approaches. 

\begin{figure}[!ht]
\centering
\includegraphics[width=1.0\columnwidth]{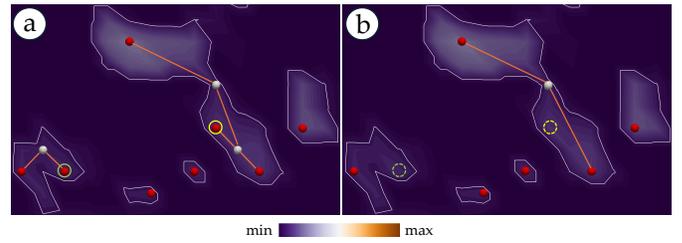}
\vspace{-6mm}
\caption{Left: a set of cloud objects enclosed by white contours; each contains a subtree of the global merge tree. Local maxima (a.k.a.,~anchor points) are in red, and saddles are in white. Right: simplifying subtrees by removing the highlighted anchor points (inside green or yellow circles) and their pairing saddles.}
\label{fig:intra-cloud-simplification-appendix}
\vspace{-6mm}
\end{figure}

\para{Intra-cloud anchor point simplification.}
When there are too many anchor points for a given cloud object, we reduce the number of anchor points for computational efficiency. In particular, we remove anchor points based on the area of their topological zones. 
For example, two of the five cloud objects
in~\cref{fig:intra-cloud-simplification-appendix}(a) contain multiple local maxima as anchor points (in red). 
We highlight the local maxima with green and yellow circles, respectively, where the associated topological zones of their adjacent edges fall below a specified threshold. 
As we remove these local maxima and their pairing saddles, we obtain the simplified subtrees inside each cloud object; 
see~\cref{fig:intra-cloud-simplification-appendix}(b). 
Subsequently, both cloud objects now contain one less anchor point.

Anchor point simplification can be prone to instabilities. Although persistence diagrams remain stable under Wasserstein and Bottleneck distances, the critical points involved in the pairings themselves may not exhibit such stability. 
For instance, in~\cref{fig:intra-cloud-simplification-appendix} (a), the lower-left component features a persistence pairing between a saddle point (shown in white) and a local maximum (in red, marked with a green circle). A slight perturbation in the underlying scalar field could cause the saddle point to change its pairing partner—potentially switching to the other local maximum (anchor point) within the same component. 
In our framework, the user defines a distance threshold to constrain anchor point movement. Anchor points from consecutive time steps are not matched if their separation exceeds this threshold. This approach helps reduce, though not entirely eliminate, the instability problem.

The threshold for intra-cloud simplification is determined based on the merge tree size and the available computational resources. Our framework computes the pFGW distance, which requires $O(n^2)$ space and $O(n^3)$ time, where $n$ denotes the (maximum) size of the merge trees. However, the actual runtime is typically lower due to the use of sparse matrix multiplication. To control merge tree size, we progressively increase the threshold, eliminating anchor points located on merge tree edges whose associated topological zones fall below the threshold. This process continues until the tree size is reduced to a manageable level. In practice, we increment the topological zone area threshold by 5, starting from 0, until all merge trees contain fewer than 5000 nodes. 
 For the {\DM} dataset, this threshold is set at $30$ pixels; for the {\DL} dataset, this threshold is set at $5$ for both the morning data and the midday data.

\begin{figure}[!ht]
\centering
\includegraphics[width=0.9\columnwidth]{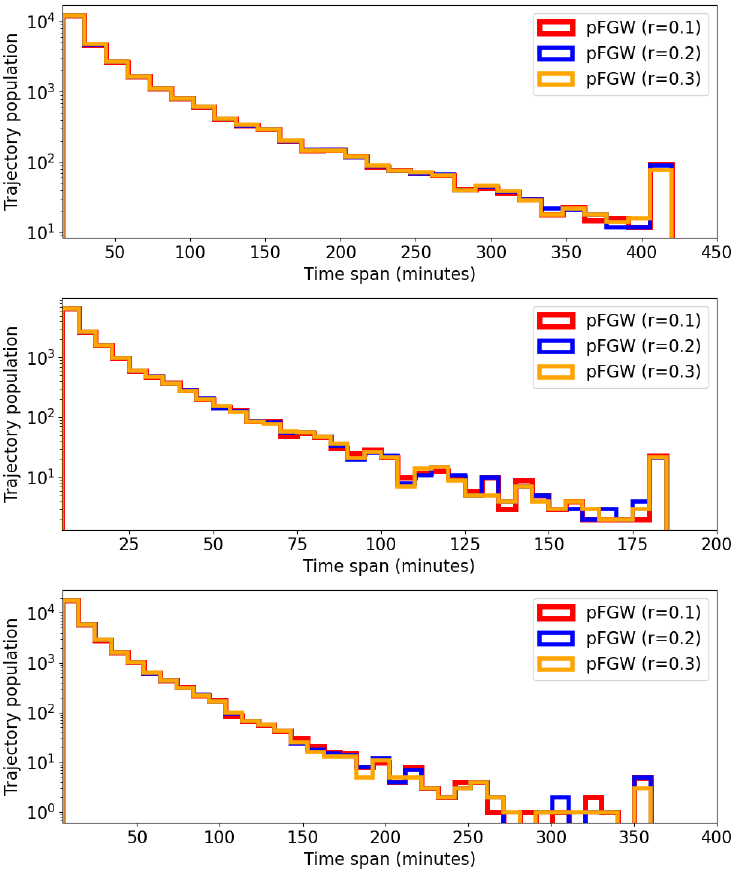}
\vspace{-2mm}
\caption{Distributions of trajectory timespans in log-scale for {\pFGW} tracking cloud systems with $r\in \{0.1, 0.2, 0.3\}$. From top to bottom: the {\DM} dataset (1st row), the {\DL} dataset during the morning (2nd row) and the midday (3rd row) period. Other experimental settings are the same as in~Sec.~6.4 and~Sec.~6.5, respectively.}
\label{fig:rvalue-timespan}
\end{figure}

\para{Strictness of matchings for cloud systems.}
In~Sec.~5.4, we discussed the strategy to compute the matching between the cloud system $X \in H_t$ and the cloud system $Y \in H_{t+1}$ using the optimal coupling matrix. 
A matching between $X$ and $Y$ is \emph{valid} if either of the two conditions is satisfied: (1) the two cloud systems $X$ and $Y$ are the mutual best match, or (2) their matching score $S_t(X, Y)$ exceeds a threshold controlled by a parameter $r$; see~Sec.~5.4 for details.
The parameter $r$ controls the strictness of the matching criteria.
When $r$ is large (e.g., \( \geq 0.5 \)), condition (2) is only met if condition (1) is also fulfilled, rendering condition (2) effectively redundant. On the other hand, when $r$ is small (e.g., \( < 0.1 \)), the matching criteria may become overly permissive, increasing the risk of mismatches despite low matching scores.  
To determine an appropriate $r$ value, we conduct experiments using the {\pFGW} framework with $r\in \{0.1, 0.2, 0.3\}$ and evaluate the resulting trajectory statistics shown in \cref{fig:rvalue-timespan}.  

\begin{figure}[!ht]
\centering
\includegraphics[width=1.0\columnwidth]{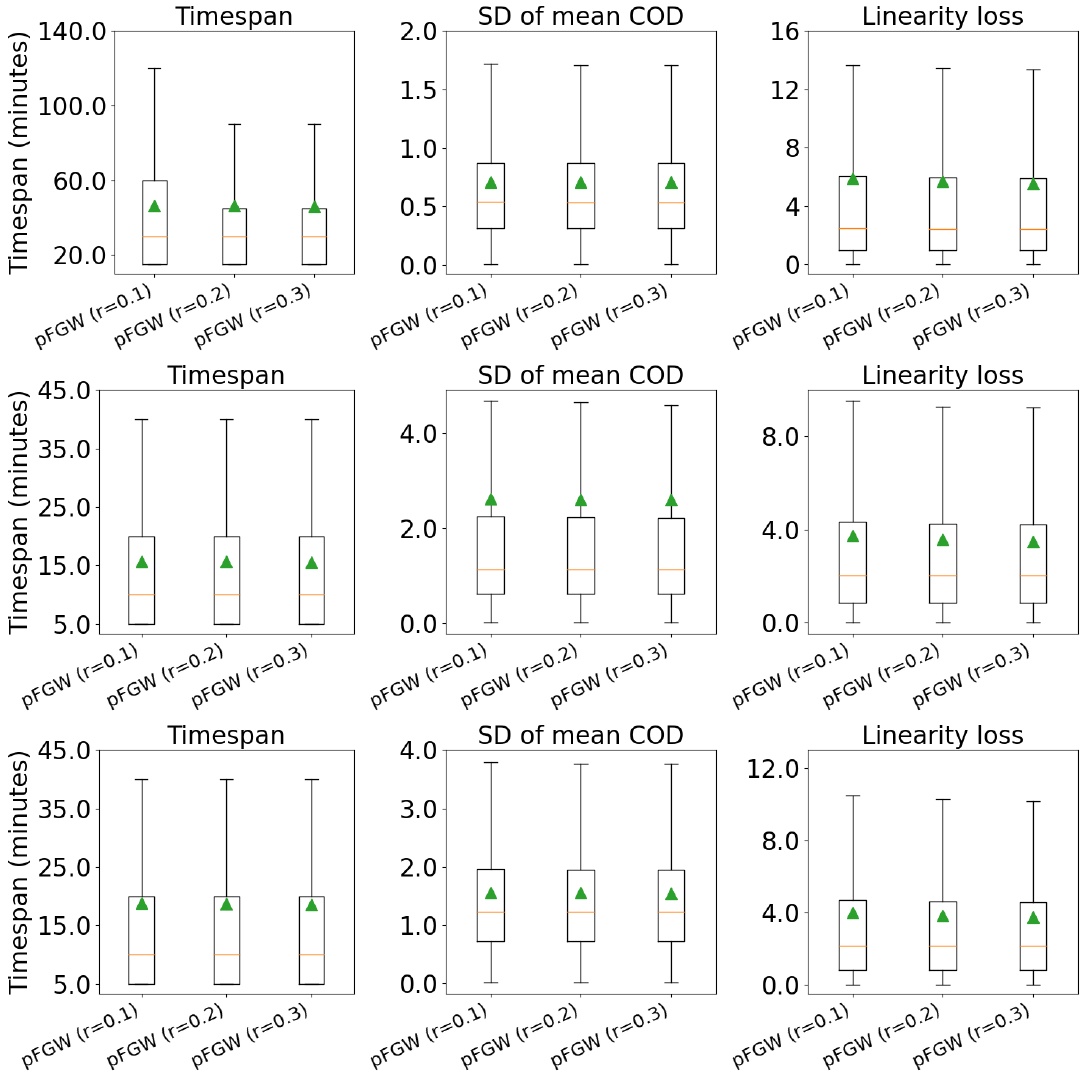}
\vspace{-4mm}
\caption{Box plots showing the median (orange line), mean (green triangle), and interquartile range (box boundary) of the distribution for three topology-based tracking methods. From top to bottom: the {\DM} dataset (1st row), the {\DL} dataset during the morning period (2nd row) and during the midday (3rd row).}
\label{fig:rvalue-statistics}
\vspace{-4mm}
\end{figure}

We first examine the trajectory timespan distributions with different choices of $r$ using the histograms in~\cref{fig:rvalue-timespan} and the box plots in~\cref{fig:rvalue-statistics} 1st column, respectively.
For the {\DM} dataset, the histograms of timespan distributions in the 1st row of~\cref{fig:rvalue-timespan} are similar. 
We only observe minor differences for trajectories with a timespan of more than $300$ minutes. 
However, in the box plots, the trajectories with $r=0.1$ have a higher interquartile range of the timespan than the ones with $r=0.2$ and $r=0.3$; see~\cref{fig:rvalue-statistics} 1st row, 1st column. 

For the {\DL} dataset, the histograms reveal more noticeable differences in the distributions of trajectory timespans; see~\cref{fig:rvalue-timespan} 2nd and 3rd rows.
For example, when $r=0.3$, the tracking results contain fewer long-lived trajectories during the midday, indicating that $r=0.3$ might be too high for the {\DL} dataset.  
Nonetheless, the overall statistics shown in the box plots of the timespan distributions for the {\DL} dataset (\cref{fig:rvalue-statistics}, second and third rows, first column) remain largely similar. Changing the parameter $r$ within the $[0.1, 0.3]$ range does not change the  timespan distribution significantly. 

We analyze the standard deviation of {\CODm} and the linearity loss for trajectories that last for at least 45 minutes (three timesteps) in the {\DM} dataset and 15 minutes (three timesteps) in the {\DL} dataset, respectively.
For both datasets, both metrics show similar distributions across all three experiments with different $r$ values; see~\cref{fig:rvalue-statistics} 2nd and 3rd columns.
For example, in the {\DM} dataset, the average standard deviation of {\CODm} (represented by green triangles in the second column of box plots in~\cref{fig:rvalue-statistics}) with 
$r=0.1$ is only $0.51\%$ higher than that with $r = 0.3$. 
Similarly, the mean trajectory linearity loss with $r=0.1$ is only $6.05\%$ higher than that with $r=0.3$. 

Based on the above statistics, we argue that the {\pFGW} tracking result is not very sensitive to the threshold parameter $r$ within the $[0.1, 0.3]$ range. 
To highlight the advantage of {\pFGW} in preserving the trajectory timespan, we select \( r = 0.1 \) for the experiments in Sec.~6.

\subsection{Parameter Configuration for Comparative Analysis}
\label{sec:parameter-Configuration}
For our comparative analysis, we report parameter configurations for {\pFGW}, {\tobac}, and {\PyFT} respectively. 

We start with parameters that are shared by all three methods. 
For all datasets, we use 8-way connectivity to search for neighboring pixels when computing superlevel set components. 
For the {\DM} dataset, we choose $2.0$ as the single COD value threshold to obtain superlevel set components; we use area-based simplification and remove the cloud objects below $10$ pixels. 
For the {\DL} dataset, we choose $9.0$ as the COD value threshold for the morning period (06:00-09:00 UTC) and $10.0$ as the threshold for the midday period (09:05-15:00 UTC); we do not simplify any cloud objects by area size. 

{\pFGW} uses the following parameters. We use $\alpha=0.2$ for {\DL} and $\alpha=0.4$ for {\DM} to balance the weight between intrinsic and extrinsic information. We adopt $30m/s$ for the {\DM} dataset and $40m/s$ for the {\DL} dataset as the maximum average speed parameter. We search the parameter $m$ within the range of $[0.6, 0.9]$ and choose the highest one where the maximum distance between matched anchor points is below the distance threshold derived from the maximum average speed.

We use the following additional parameters when running the {\PyFT} framework:
We use superlevel set components as cloud objects.
For both datasets, the maximum number of objects for a cloud object to correspond to in the next time step is $10$. The overlap percentage threshold is $30\%$. The maximum number of clouds in the domain is $3000$ for every time step. When a cloud object splits, the minimum region overlap size for the main trajectory to be assigned to is $3$ pixels for {\DM} and $1$ pixel for {\DL}. 

Our experiments with {\tobac} use the following additional parameters. 
During cloud detection, {\tobac} uses the barycenter of the bounding box of detected features as feature points of cloud objects. 
The barycenter is weighted by the COD value minus the superlevel set threshold. 
This barycenter strategy is labeled ``weighted\_diff'' in the {\tobac} tool.
When tracking feature points, {\tobac} uses the ``predict'' method, which predicts the translation direction of the feature based on its previous trajectory. 
Besides, {\tobac} uses adaptive search for the maximum size of a feature point set to locate the matching feature in the next time step. The initial value of this parameter is $20$, and decays at a speed of $0.9$ until it reaches $5$.
For the {\DM} dataset, the maximum average speed of clouds between adjacent snapshots is $20m/s$. For the {\DL} dataset, it is $40m/s$.

\subsection{Limitations of Comparative Analysis}
There is a limitation in our comparative analysis: {\tobac} employs a multi-threshold cloud detection process while both {\pFGW} and {\PyFT} uses a single threshold for cloud detection. This multi-threshold detection process can find superlevel sets at different thresholds and summarize all the detected cloud objects. A typical use of this process is to break down a large cloud object into several objects at a higher threshold. For a fair comparison, we use the same threshold for cloud detection for all three methods. 

%% file: sec-topotools-comparison.tex
 \begin{figure*}[!ht]
\centering
\vspace{-2mm}
\includegraphics[width=1.8\columnwidth]{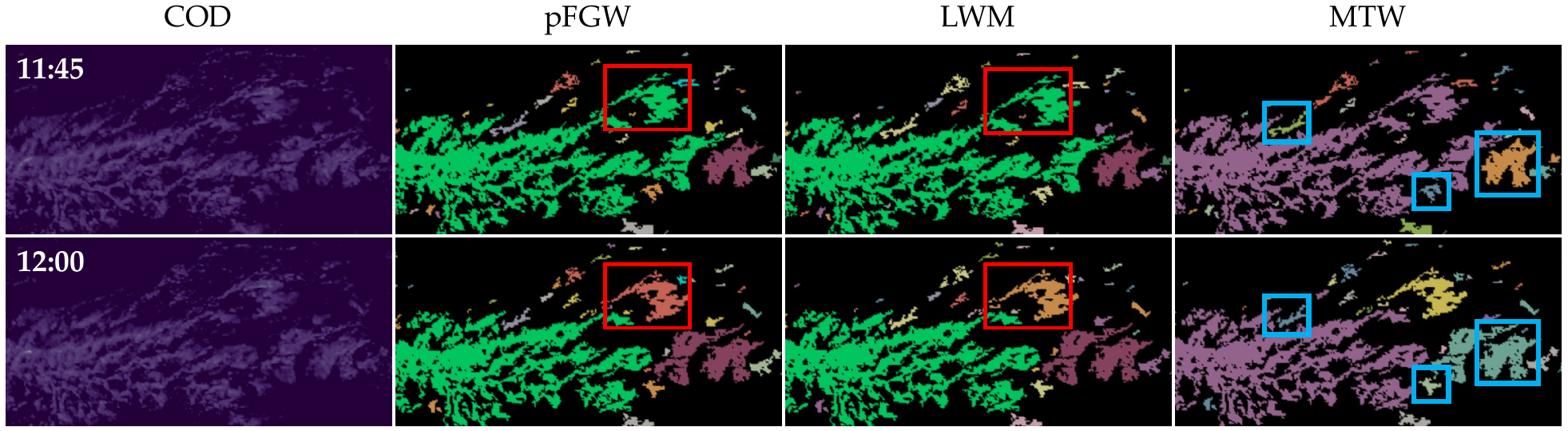}
\vspace{-2mm}
\caption{Tracking results of a region in the {\DM} dataset on Aug 1, 2023. From left to right: visualizations of COD fields, tracking results for {\pFGW}, {\LWM}, and {\MTW}, respectively. 
Red boxes highlight the area where {\pFGW} and {\LWM} have different tracking results for a cloud splitting event (see~\cref{fig:topo-marine-case-zoomin} for zoom-in views). Cyan boxes highlight multiple areas where {\MTW} generate incorrect correspondences for cloud systems.
}
\label{fig:topo-marine-case-study}
\vspace{-6mm}
\end{figure*}

\section{Comparison with Topology-Based Tracking Tools}
\label{sec:supplement-topology-tools}

In this section, we provide additional details for comparing the three topology-based tracking tools: {\pFGW}, {\LWM}, and {\MTW}; see~Sec.~6.6 for an overview. 
We start with the experimental settings for {\LWM} and {\MTW} in~\cref{sec:supplement-topotool-experiment}.
Next, we provide a qualitative evaluation using the {\DM} dataset in~\cref{sec:supplement-topotool-marine}, followed by a comparison using the {\DL} dataset in~\cref{sec:supplement-topotool-land}. 

\subsection{Experimental Settings}
\label{sec:supplement-topotool-experiment}
\para{Preprocessing.}
We follow the same process in~Sec.~5.1 and~Sec.~5.2 to obtain the simplified merge trees of the COD fields. 
{\MTW} directly uses the simplified merge trees as the input. For {\LWM}, we compute the persistence-based branch decomposition to obtain the persistence diagrams as the input. 

\para{Implementation.}
Both {\MTW} and {\LWM} are implemented as built-in modules of the Topology Toolkit (TTK)~\cite{TiernyFavelierLevine2018}.
We use the TTK module for {\MTW} computation.
However, the TTK module for {\LWM} only accepts simplified scalar fields as the input, while we can only obtain simplified merge trees (or persistence diagrams). Therefore, we implement the {\LWM} approach in-house by modifying the code from the TTK module.

\para{Parameter settings.}
For {\pFGW}, we use an Euclidean distance threshold at $28km$, which corresponds to an average speed of $\approx 30m/s$; see~\cref{sec:parameters-tracking} for a discussion on parameter configurations. 
The parameter settings of {\LWM} focus on the weight balancing the Wasserstein distance between persistence pairs and the Euclidean distance between critical points. 
As described in~Sec.~6.6, we normalize the range of the two distances and set the weight of the Euclidean distance to \( 1.0 \). We then gradually increase the weight of the Wasserstein distance, denoted as \( \beta \), from \( 0 \) to assess the impact of persistence information. 
For {\MTW}, the parameter $\varepsilon_1$ is a threshold to determine whether the saddles for two branches on the tree can be swapped, whereas $\varepsilon_2$ and $\varepsilon_3$ are used to control the persistence scaling during the transformation between branch decomposition trees and merge trees. Due to the high complexity and uncertainty~\cite{Platnick2017} of the COD field, we have no prior knowledge about persistence to tune these parameters. Instead, we follow the recommended parameter setting from its original work~\cite{PontVidalDelon2021}. 

\para{Postprocessing.}
We compute the cloud system trajectories for {\LWM} and {\MTW} using a postprocessing pipeline similar to that of {\pFGW}. First, we identify anchor points within the cloud system and determine matching scores based on the number of anchor point matchings. Then, we apply the algorithm described in~Sec.~5.4 to extract the main trajectories of the cloud systems. 
We compare the cloud system tracking performance using the statistics described in~Sec.~6.3.

\subsection{Case Study: Marine Cloud Dataset}
\label{sec:supplement-topotool-marine}
\para{Cloud tracking.}
In~Sec.~6.6, we have evaluated the anchor point matching results and the statistics of cloud system trajectories using the three metrics in~Sec.~6.3.
In this section, we examine the visualizations of tracking results using the data from Aug 1, 2023.
We check the transition from 11:45 to 12:00 UTC. 

During the transition, a splitting event occurs, as highlighted by the red boxes in~\cref{fig:topo-marine-case-study}. All three methods have detected the splitting event. However, {\pFGW} and {\LWM} demonstrate different cloud system trajectories for this event. 

\begin{figure}[!ht]
\centering
\includegraphics[width=1.0\columnwidth]{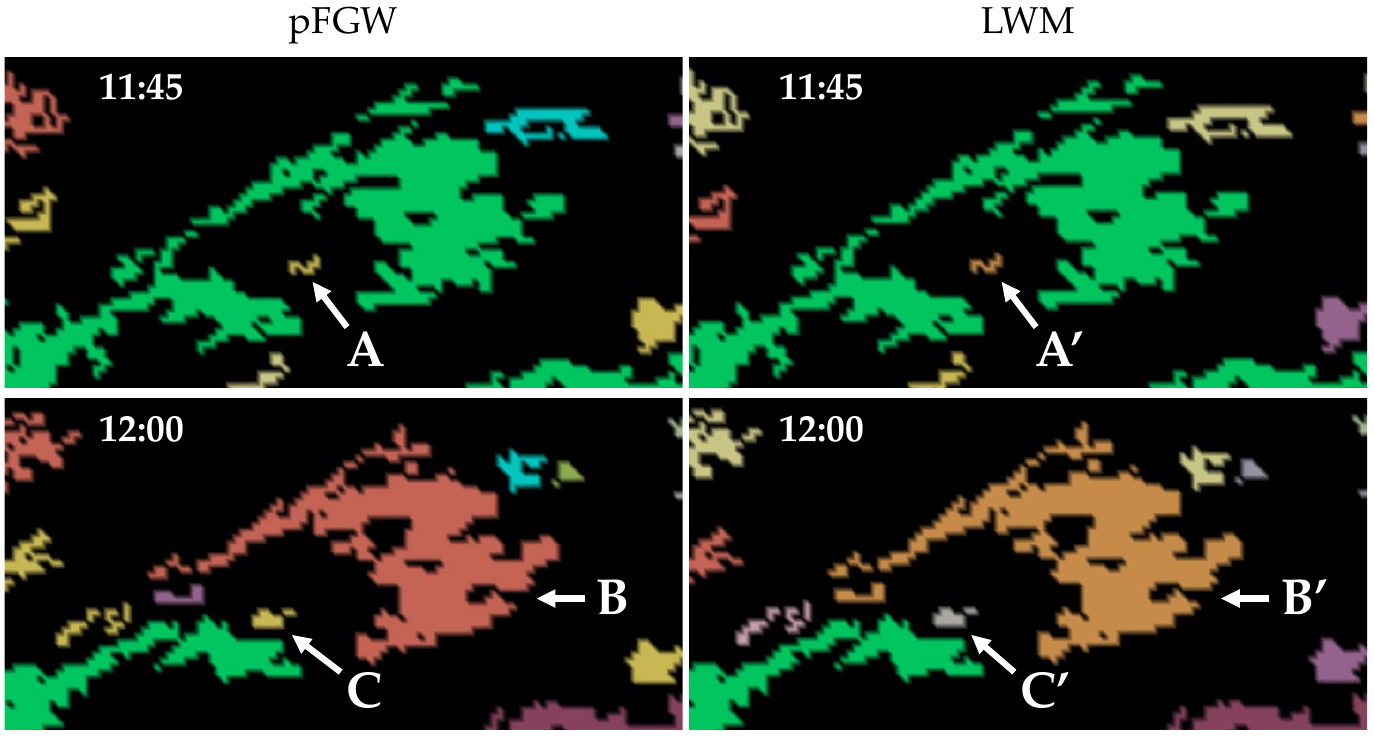}
\vspace{-4mm}
\caption{{\DM} dataset: Zoom-in views for~\cref{fig:topo-marine-case-study} red boxes to demonstrate the trajectory differences between {\pFGW} and {\LWM}. {\pFGW} matches cloud system $A$ to $C$, whereas {\LWM} matches $A'$ to $B'$.}
\label{fig:topo-marine-case-zoomin}
\vspace{-3mm}
\end{figure}

A closer examination in the zoomed-in view in~\cref{fig:topo-marine-case-zoomin} reveals that in the {\pFGW} results, cloud system $A$ at 11:45 is matched to cloud system $C$ at 12:00, while cloud system $B$ is identified as newly formed after splitting from a larger green cloud system. This matching is reasonable given the geometric proximity of $A$ and $C$.

In contrast, {\LWM} matches cloud system $A'$ at 11:45 to $B'$ at 12:00 and considers cloud system $C'$ as newly formed. This results in a significant jump in the centroid location from  $A'$ to $B'$, increasing the linearity loss of the trajectory. 
Moreover, we do not observe substantial changes in the COD fields in the corresponding areas in~\cref{fig:topo-marine-case-study}. The matching between cloud system $A'$ and $B'$ is likely a mismatch.
Such errors in {\LWM} can occur when the cost of matching the anchor points in the cloud system $B'$ to the diagonal (representing their ``birth'') is higher than the cost of matching them to other faraway anchor points.

On the other hand, {\MTW} fails to track many cloud systems accurately. The cyan boxes in~\cref{fig:topo-marine-case-study} highlight multiple areas in which {\MTW} does not generate matchings between the two time steps. 
This result shows that {\MTW} performs the worst among the three topology-based approaches for this cloud tracking task.

\begin{figure*}[!ht]
\centering
\vspace{-2mm}
\includegraphics[width=1.8\columnwidth]{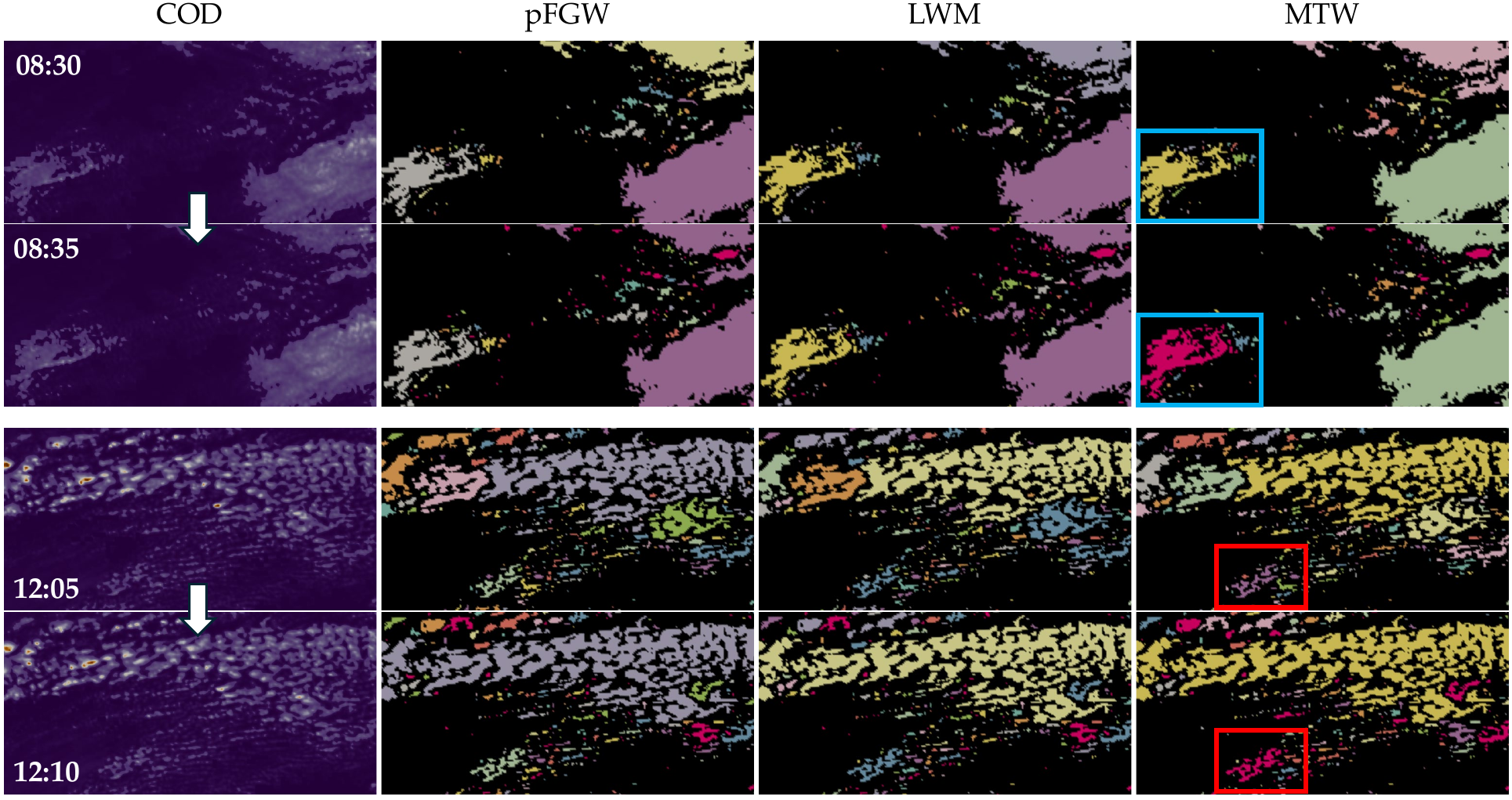}
\vspace{-2mm}
\caption{Tracking results of a region in the {\DL} dataset on May 1, 2018. From left to right: visualizations of COD fields, tracking results for {\pFGW}, {\LWM}, and {\MTW}, respectively. 
The top two rows (resp. bottom two rows) are for the transition during the morning (resp. midday) period. 
All new cloud entities in the 2nd and 4th rows are colored magenta; others are colored by correspondences. 
Cyan boxes and red boxes highlight the areas where {\MTW} fails to track cloud systems with noticeable areas in the morning and the midday, respectively.
}
\label{fig:topo-land-case-study}
\vspace{-6mm}
\end{figure*}

\begin{figure}[!ht]
\centering
\includegraphics[width=0.9\columnwidth]{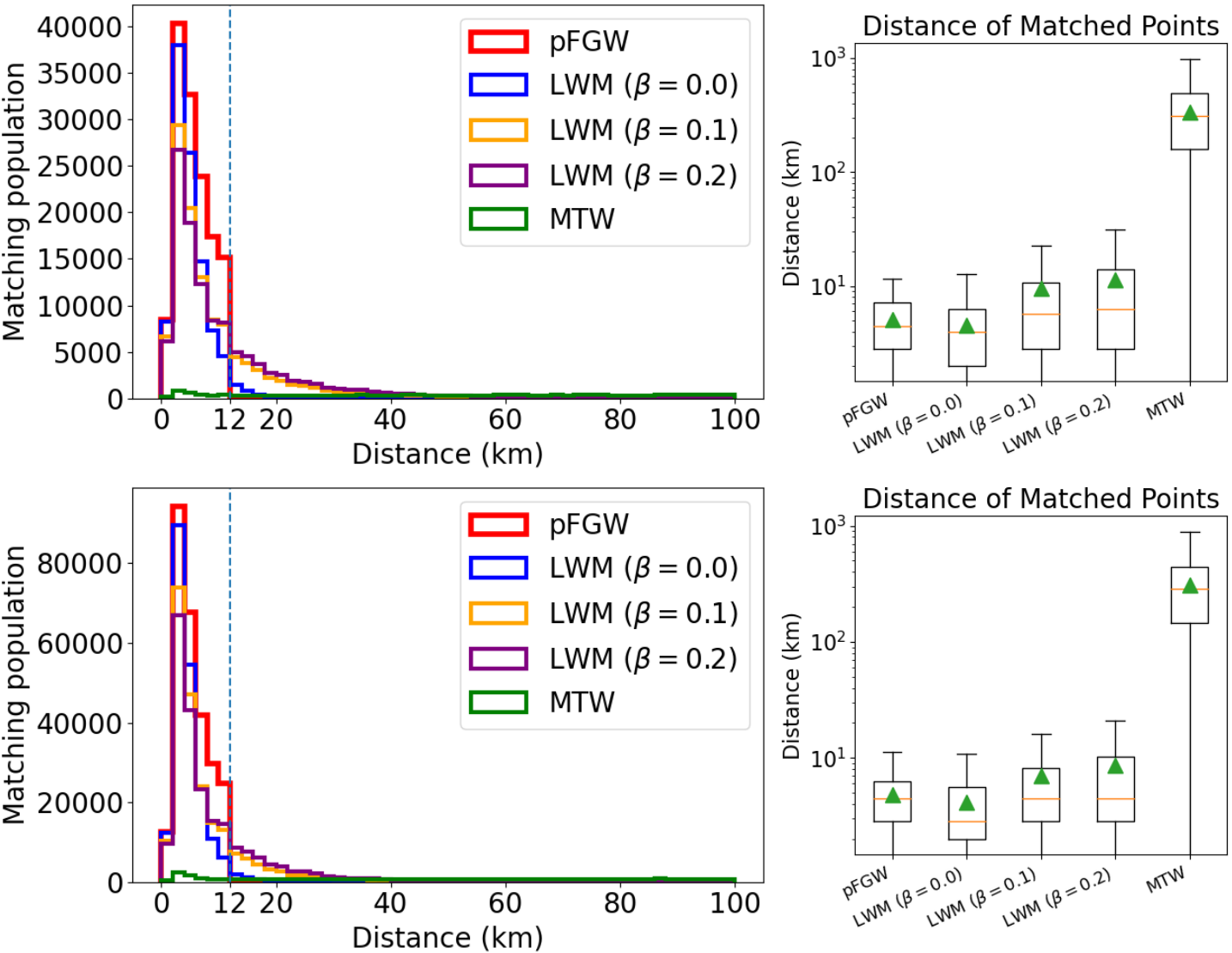}
\vspace{-2mm}
\caption{{\DL} dataset: histograms (left) and box plots (right) for distributions of the Euclidean distances between matched anchor points. The Euclidean distance threshold for {\pFGW} is $12$km, as highlighted by the vertical dotted line in the histograms. The top row is for the morning data, and the bottom is for the midday data.}
\label{fig:topo-land-distance}
\vspace{-3mm}
\end{figure}

\begin{figure}[!ht]
\centering
\includegraphics[width=0.95\columnwidth]{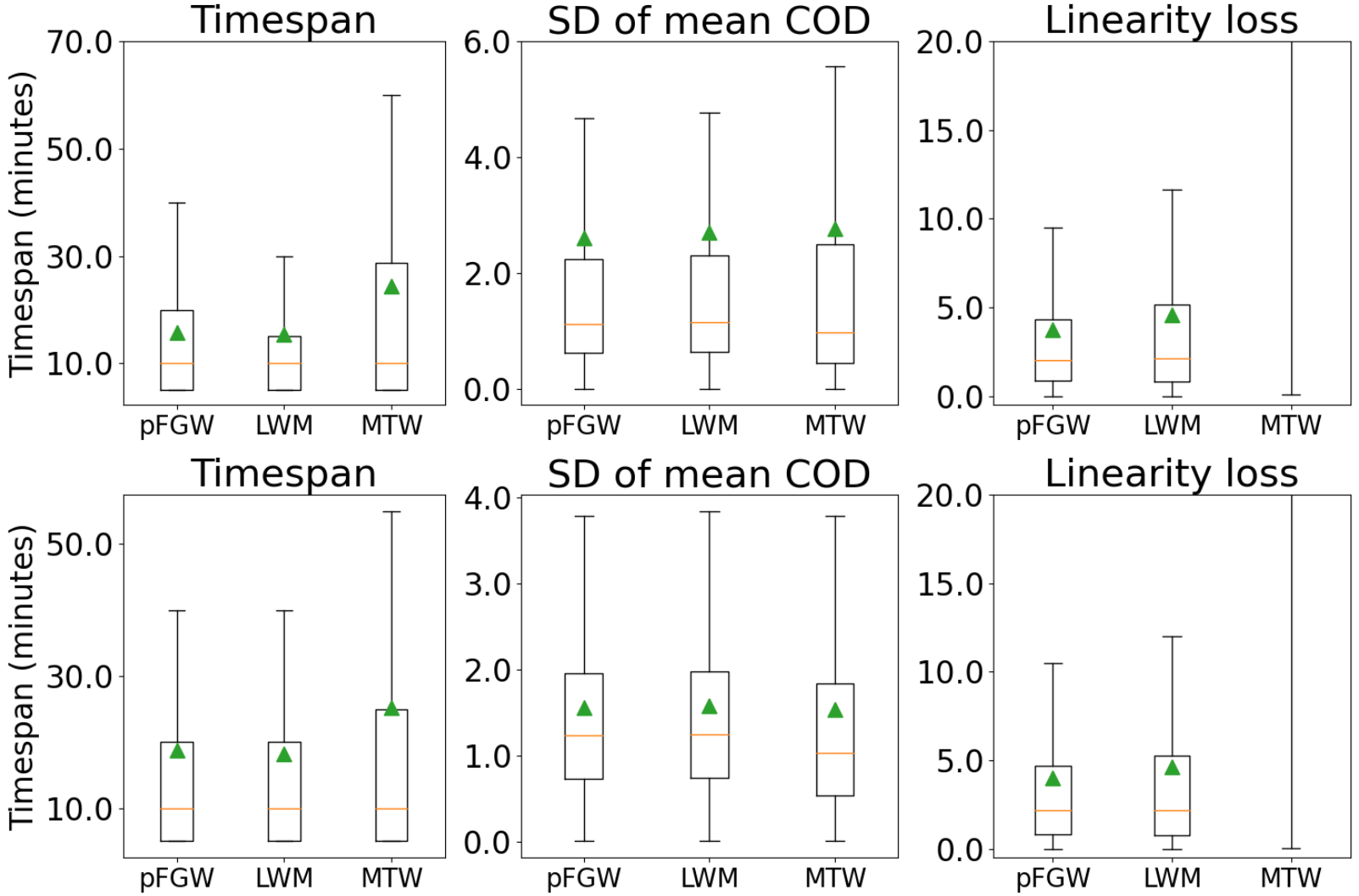}
\vspace{-2mm}
\caption{{\DL} dataset: box plots showing the median (orange line), mean (green triangle), and interquartile range (box boundary) of the distribution for three topology-based tracking methods. The top row is for the morning data, whereas the bottom is for the midday data. The boxes for the linearity loss for {\MTW} exceed the plots' upper bound.}
\label{fig:topo-land-statistics}
\vspace{-3mm}
\end{figure}

\subsection{Case Study: Land Cloud Dataset}
\label{sec:supplement-topotool-land}
We provide the experimental results and statistical evaluations for the three topology-based methods on the {\DL} dataset.

\para{Anchor point matching.}
Following the approach in~Sec.~6.6, we evaluate the effect of persistence diagram information on tracking for {\LWM} using $\beta \in \{0.0, 0.1, 0.2\}$.
The Euclidean distance distributions between matched anchor points are presented in~\cref{fig:topo-land-distance}. 
For {\pFGW}, we include all matchings with nonzero probability in the distributions, whereas {\LWM} and {\MTW} provide one-to-one anchor point matchings.

For both the morning and midday data, the Euclidean distance distributions of {\pFGW} and {\LWM} results are similar. In particular, for {\LWM} $(\beta=0)$, only a small fraction of matched anchor points have distances greater than $12$km; see~\cref{fig:topo-land-distance} 1st column.
In contrast, {\LWM} with $\beta=0.1$ and $\beta=0.2$ generates fewer short-distance matchings and more long-distance matchings between anchor points compared to {\LWM} with $\beta=0$. 
This observation indicates that $\beta=0$ is the optimal choice for {\LWM}. 
Increasing the weight of the Wasserstein distance in the cost function does not improve the accuracy of cloud tracking.

We notice that the mean and median Euclidean distances between matched anchor points for {\LWM} $(\beta=0)$ are lower than those of {\pFGW}; see~\cref{fig:topo-land-distance} 2nd column. This is because {\pFGW} results include more pairs of matched anchor points with Euclidean distances close to $12$km; see~\cref{fig:topo-land-distance} 1st column. 
However, {\pFGW} results may include low-probability anchor point matchings, whose influence on the cloud system tracking results is minimal. Further evaluation of cloud trajectory statistics is necessary for a comprehensive comparison.

As with the {\DM} dataset, {\MTW} performs the worst in matching nearby anchor points for the {\DL} dataset across the morning and midday. The mean and median Euclidean distances for matched anchor points using {\MTW} are significantly higher than the other two methods; see~\cref{fig:topo-land-distance} 2nd column. 

\para{Statistical evaluation.}
\cref{fig:topo-land-statistics} presents the distribution of evaluated metrics for the three topology-based approaches. 
For {\LWM}, we fix $\beta=0$ because it performs the best for anchor point matching among all choices of $\beta$.

We first examine the trajectory timespan distribution. {\MTW} results exhibit the highest mean trajectory timespan for both the morning and the midday data. Between the other two methods, {\pFGW} achieves a slightly higher mean trajectory timespan than {\LWM} for both the morning and midday periods; see~\cref{fig:topo-land-statistics} 1st column. 

Next, we check the standard deviation of the mean COD for trajectories that last for at least 15 minutes (three time steps).
All three approaches have comparable distributions for this metric across the morning and midday; see~\cref{fig:topo-land-statistics} 2nd column. 

For the trajectory linearity loss, {\MTW} trajectories exhibit significantly higher loss than those from {\pFGW} and {\LWM}, indicating frequent mismatches between distant cloud systems (\cref{fig:topo-land-statistics} 3rd column). 
Comparing {\pFGW} against {\LWM}, {\pFGW} trajectories have a lower mean linearity loss and interquartile range than {\LWM} (\cref{fig:topo-land-statistics} 3rd column), indicating better preservation of linearity. 
Overall, {\pFGW} demonstrates the most reliable performance in maintaining the linearity of cloud system trajectories among the three topology-based approaches.

\para{Cloud tracking.}
Lastly, we compare the cloud tracking results on the {\DL} dataset using visualizations for our case study.
We use the same set of data in~Sec.~6.5 for evaluation. That is, we examine the time transition from 08:30 to 08:35 UTC for the morning data and from 12:05 to 12:10 UTC for the midday.
New cloud entities at 8:35 UTC and 12:10 UTC are colored in magenta, which we use to evaluate the ability of each method to maintain consistent cloud system tracking.

The tracking results are demonstrated in~\cref{fig:topo-land-case-study}.
During the morning transition (1st and 2nd rows),  small cloud systems emerge near the center of the COD field. 
For this transition, {\pFGW} and {\LWM} generate similar results in tracking these cloud systems; see the 2nd and 3rd columns of the first two rows. 
{\MTW}, on the other hand, identifies many small cloud systems as tracked from the previous time step in the morning period. 
For example, in the 2nd row 4th column of~\cref{fig:topo-land-case-study}, many small cloud systems in the center of the image are not colored in magenta, indicating that they are tracked from other systems at the previous time step.
This explains why the mean and median timespans for {\MTW} trajectories are high.
However, the tracking quality of the {\MTW} results is undesirable.
We cannot find many color correspondences for these small cloud systems, see the 1st and 2nd rows of the 4th column in~\cref{fig:topo-land-case-study}. The cloud system matching between the two time steps looks random.
In other words, {\MTW} trajectories are unlikely to reflect the actual cloud system movement.

In the {\DL} dataset, the situation of random matching for {\MTW} occurs because the anchor points of the small cloud systems tend to have their COD values (as well as the persistence of their pairs) similar to the cloud detection threshold.
{\MTW}, relying on the persistence information for critical point matching, tends to match anchor points with similar persistence information. 
Consequently, {\MTW} tends to match these small cloud systems without considering the geometric proximity.

The tracking results for the midday transition (cf.~\cref{fig:topo-land-case-study} 3rd and 4th rows) yield similar findings. Both {\pFGW} and {\LWM} produce visually comparable results, whereas {\MTW} fails to track cloud systems correctly. 
For example, {\MTW} loses the trajectory for two prominent cloud systems in~\cref{fig:topo-land-case-study} cyan boxes during the morning transition (see 1st and 2nd rows, 4th column) and in~\cref{fig:topo-land-case-study} red boxes during the midday transition (see 3rd and 4th rows, 4th column), respectively. This further demonstrates the limitations of {\MTW} in tracking cloud systems.